\documentclass[letterpaper,twocolumn,10pt]{article}
\usepackage{usenix2019_v3}

\usepackage{tikz}
\usepackage{amsmath}

\usepackage{filecontents}

\usepackage{lipsum}
\usepackage{enumitem}
\usepackage{stfloats}
\usepackage{tcolorbox}
\usepackage[ruled,linesnumbered]{algorithm2e}
\usepackage{multirow}
\usepackage{graphicx}
\usepackage{float}
\usepackage{booktabs}
\usepackage{subfigure} 
\usepackage{xcolor}

\usepackage{makecell}  %

\usepackage{enumitem}

\usepackage{pifont}     %
\usepackage{circledsteps}   %

\usepackage{tcolorbox}
\tcbuselibrary{skins, breakable}

\usepackage{setspace}

\newcommand{\circnum}[1]{\raisebox{.5pt}{\textcircled{\raisebox{-.9pt} {\scriptsize #1}}}}

\newcommand{\myparagraph}[1]{%
    \noindent\textbf{#1}  %
}

\newcommand{\speedup}{8.65}

\begin{document}

\date{}

\newcommand{\methodName}{NeTestLLM}

\title{Automated Network Protocol Testing with LLM Agents}

\author{
Yunze Wei$^{1}$,
Kaiwen Chi$^{1}$,
Shibo Du$^{1}$,
Jianyu Wang$^{1}$,
Zhangzhong Liu$^{1}$,
Yawen Wang$^{1}$,\\
Zhanyou Li$^{2}$,
Congcong Miao$^{3}$,
Xiaohui Xie$^{1}$\footnotemark[1],
Yong Cui$^{1}$\footnotemark[1]\\[4pt]
$^{1}$Tsinghua University \quad
$^{2}$Beijing Xinertel Technology Co., Ltd. \quad
$^{3}$Tencent
}

\maketitle

\renewcommand{\thefootnote}{\fnsymbol{footnote}}
\footnotetext[1]{Corresponding authors: Xiaohui Xie (xiexiaohui@tsinghua.edu.cn), Yong Cui (cuiyong@tsinghua.edu.cn).}

\begin{abstract}

Network protocol testing is fundamental for modern network infrastructure.
However, traditional network protocol testing methods are labor-intensive and error-prone, requiring manual interpretation of specifications, test case design, and translation into executable artifacts, typically demanding one person-day of effort per test case.
Existing model-based approaches provide partial automation but still involve substantial manual modeling and expert intervention, leading to high costs and limited adaptability to diverse and evolving protocols.
In this paper, we propose a first-of-its-kind system called \methodName{} that takes advantage of multi-agent Large Language Models (LLMs) for end-to-end automated network protocol testing. 
\methodName{} employs hierarchical protocol understanding to capture complex specifications, iterative test case generation to improve coverage, a task-specific workflow for executable artifact generation, and runtime feedback analysis for debugging and refinement.
\methodName{} has been deployed in a production environment for several months, receiving positive feedback from domain experts. %
In experiments, \methodName{} generated 4,632 test cases for OSPF, RIP, and BGP, covering 41 historical FRRouting bugs compared to 11 by current national standards. 
The process of generating executable artifacts also improves testing efficiency by a factor of \speedup{}x compared to manual methods. %
\methodName{} provides the first practical LLM-powered solution for automated end-to-end testing of heterogeneous network protocols.

\end{abstract}

\section{Introduction}

Network protocol testing is crucial for modern communication infrastructure. 
It ensures interoperability, reliability, and security across routers, switches, firewalls, and other network devices. 
Protocol standards are continuously updated by standardization organizations such as the IETF, IEEE, and by industry alliances.
Emerging domains such as satellite internet~\cite{lai2025leocc, liu2024democratizing} and internet of things~\cite{zhang2025towards} further accelerate the deployment of new protocols, whose implementations require rigorous validation prior to operational use.
As a result, protocol testing is required throughout the entire device lifecycle, from design and development to deployment and operation.

Conventional network protocol testing is predominantly manual.
The process requires engineers to design test cases from protocol specifications, convert them into executable artifacts such as tester scripts and the device under test (DUT) configurations, and iteratively refine them in testing environment.
This process necessitates close coordination among service purchasers, tester providers, and DUT vendors.
According to feedback from production network operations, designing and implementing a single test case typically requires one person-day of effort.
It is time-consuming, labor-intensive, and error-prone, while offering limited consistency across projects and poor adaptability to the rapid evolution of standards and devices.
Recent model-based approaches, such as SCALE~\cite{kakarla2022scale} and MESSI~\cite{singha2024messi}, introduce partial automation but still rely on costly manual modeling and lack the flexibility to accommodate diverse and evolving requirements.

Recent advances in large language models (LLMs) open up new opportunities for revolutionizing this workflow.
Existing studies have already applied LLMs to various areas of software testing~\cite{chen2024chatunitest, wang2024hits, deng2024pentestgpt, meng2024large}.
However, these approaches cannot be directly migrated to the domain of network protocol testing.
Unlike software testing, where test cases are often executable, network protocol test cases are usually written in natural language and must be transformed into executable artifacts before execution.
In this context, we argue that LLMs hold significant potential in parsing specification documents, generating test cases, producing executable artifacts, and analyzing execution logs, 
highlighting their potential for automated network protocol testing approaches.

In this paper, we propose a first-of-its-kind system called \methodName{} that takes advantage of multi-agent LLMs for end-to-end automated network protocol testing (\S~\ref{sec:design-workflow}). 
\methodName{} employs hierarchical protocol understanding to capture complex specifications, iterative test case generation to improve coverage, task-specific workflow for executable artifact generation, and runtime feedback analysis for debugging and refinement.
Our overarching goal is to minimize human intervention while maintaining high reliability. %
\methodName{} achieves its goal through the following design aspects:

\myparagraph{Understanding protocol specifications (\S~\ref{subsec:rfc-understanding}).}
Protocol specifications are complex, making it challenging for LLMs to capture both high-level semantics and fine-grained details.
To address this, we design a hierarchical protocol understanding pipeline that combines high-level function modeling with low-level module modeling. %

\myparagraph{Evaluating and refining test cases (\S~\ref{subsection:testcase-gen}).}  %
Test cases expressed in natural language lack standardized quality metrics, making it difficult to assess their reliability and coverage.  
To address this, we design a semi-quantitative evaluation mechanism based on key section coverage and semantic coverage, guiding iterative refinement of generated test cases.  

\myparagraph{Generating executable artifacts (\S~\ref{section:artifact-gen}).}  %
Generating executable artifacts such as tester scripts and DUT configurations requires private domain knowledge and vendor-specific expertise that LLMs do not inherently possess. 
We enhance LLM agents with domain knowledge base and standard operating procedures (SOPs) to bridge this gap. %

\myparagraph{Analyzing unified execution logs (\S~\ref{sec:runtime-feedback}).}
Execution logs may simultaneously reflect diverse sources of errors, including DUT implementation bugs, misconfigurations, tester script errors, and flawed test cases, making root cause analysis difficult. 
We address this with a hierarchical feedback loop that iteratively refines executable artifacts and test cases, isolating errors to their likely sources.

\methodName{} has been deployed in the production environment for several months, receiving positive feedback from domain experts. 
In the evaluation (\S~\ref{sec:evaluation}), our test case generation module produced 4,632 test cases for three mainstream routing protocols, achieving substantially higher coverage than existing national standards. %
Our generated test cases covered 41 FRRouting~\cite{frrouting:frr} historical bugs, compared to 11 by the current national standards.
The executable artifact generation module also improved efficiency by a factor of 8.65x over the typical manual process. %
In an expert user study, the generated test cases and executable artifacts achieve average scores of 8.40 and 7.24 out of 10 respectively, indicating they are \textit{very helpful}. 
\methodName{} effectively automates network protocol testing, reducing human effort while enhancing coverage.

\myparagraph{Contributions:} We make the following contributions:

\begin{itemize}[itemsep=-1ex, topsep=0.1em]
    \item The first end-to-end network protocol testing framework leveraging multi-agent LLMs. %
    \item A hierarchical protocol understanding pipeline, and an iterative test case generation and verification method. %
    \item A task-specific workflow for generating executable artifacts, and a runtime feedback analysis mechanism. %
    \item System implementation, expert evaluation and experiments on real-world datasets. %
\end{itemize}

This work \textbf{does not} raise any ethical issues.

\section{Background and Motivation}\label{sec:background}

\begin{figure}[t]
    \centering
    \includegraphics[width=0.86\linewidth]{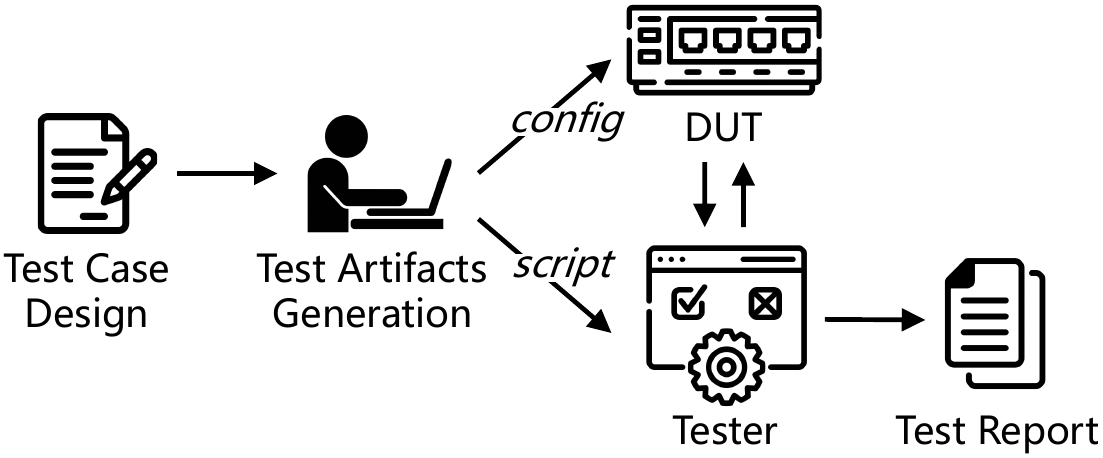}
    \vspace{-1.5ex}
    \caption{Network protocol testing workflow.}
    \label{fig:basic_workflow}
    \vspace{-1ex} 
\end{figure}

\subsection{Network Protocol Testing}
\label{sec:device-testing}

Network protocol testing is a critical process throughout the lifecycle of network devices, verifying compliance with functional, performance, security, and interoperability requirements. 
It proactively identifies potential failures, enforces standards and cross-vendor interoperability, and mitigates security risks to maintain network stability.
Conducting this process requires engineers to design test cases based on protocol specifications, convert them into executable artifacts for both the tester and the device under test (DUT), and iteratively refine them within the testing environment (see Figure~\ref{fig:basic_workflow}).

There are several core categories of network protocol testing, such as protocol conformance, functional verification, performance evaluation, security assessment, and interoperability validation.
A typical testing setup involves a \textit{tester} and a \textit{DUT}. 
The tester is a dedicated instrument that generates network traffic, receives and analyzes responses, and supports a wide range of protocols. 
It can emulate the behavior of various network devices to interact with the DUT. 
The DUT refers to the target device being tested, such as a router, switch, or firewall. 
Executing a test case usually involves both code scripts running on the tester and corresponding configuration files applied to the DUT, ensuring that traffic patterns and device behaviors are jointly validated.

\subsection{Motivation}\label{subsec:motivation}

Network protocol testing plays a critical role throughout the lifecycle of network devices. 
It typically takes place in the following key scenarios:
(1) during device development, led by vendors;
(2) during procurement and acceptance, where customers conduct proof-of-concept (PoC) and formal acceptance testing.

\myparagraph{Example scenario 1: Device development testing.}\label{subsubsec:example_1}
This stage is conducted by vendors, aiming to validate hardware soundness and software completeness. 
Testing in this phase facilitates the early identification of design flaws and functional vulnerabilities.%
Development testing typically combines white-box and gray-box approaches.
White-box testing leverages full internal knowledge to carry out in-depth verification, while grey-box testing simulates real-world interactions and balance coverage with efficiency.

\begin{figure}
    \centering
    \includegraphics[width=0.93\linewidth]{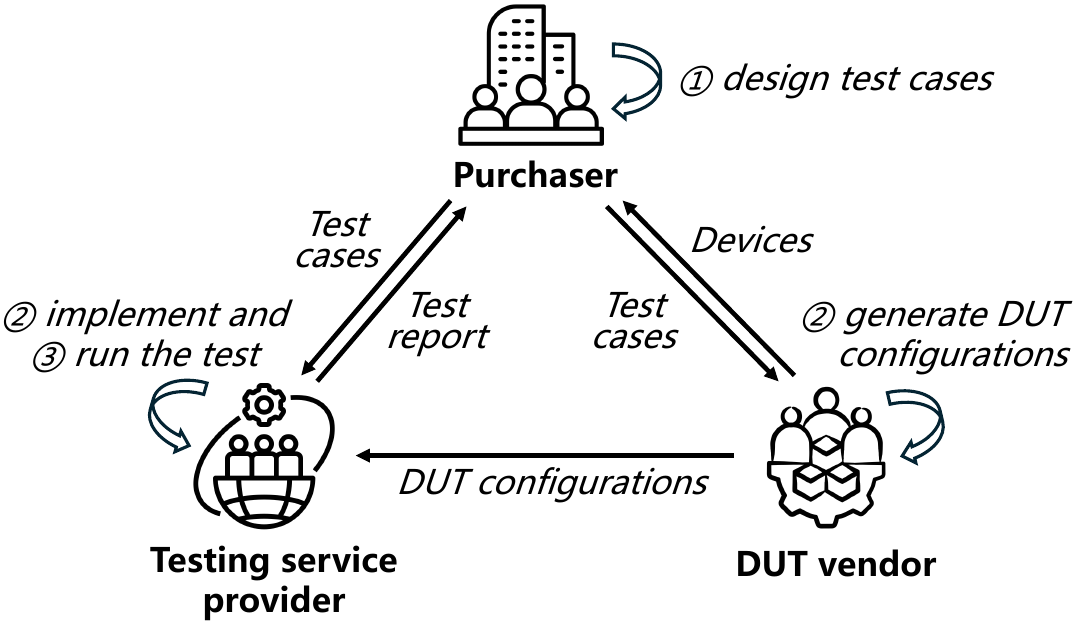}
    \caption{A typical scenario of network protocol testing involving multiple stakeholders.}
    \label{fig:motivation}
\end{figure}

\myparagraph{Example scenario 2: Device acceptance testing.}\label{subsubsec:example_2}
Device acceptance testing is performed after manufacturing and delivery but prior to deployment. 
Its goal is to confirm that devices meet contractual specifications and business requirements, and can be integrated into the target network.
Unlike development testing, it is typically conducted as a black-box process, focusing on externally observable behaviors rather than internal design details.

\myparagraph{Limitations of current testing workflows.}
Current protocol testing workflows remain heavily manual and require close collaboration among multiple stakeholders, as illustrated in Figure~\ref{fig:motivation}.
In device acceptance testing, the \textit{purchaser} (e.g., network operators or large enterprises) designs test cases based on protocol specifications, standards and business needs.
The test cases are then delivered to the \textit{testing service provider}, who translates them into tester-specific executable scripts. %
Meanwhile, the \textit{DUT vendor} prepares the corresponding device configurations required by the test cases.
These artifacts are subsequently validated through iterative debugging in a testbed and, once finalized, executed in the dedicated test environment to generate testing reports.
Each test case typically takes at least one person-day to design and implement.
This multi-party process often incurs additional delays of days to months, making the entire workflow time-consuming, labor-intensive, and error-prone, limiting its adaptability to rapidly evolving protocols and heterogeneous device requirements.

\myparagraph{Opportunities with LLM Agents.}
Recent advances in LLM agents~\cite{hong2024metagpt, wu2024autogen} open up new opportunities to alleviate existing limitations in network protocol testing. 
LLM agents can fully leverage the underlying strengths of LLMs in natural language understanding, code generation, and knowledge transfer.
These capabilities can be harnessed to automatically parse and understand protocol specifications, generate test cases, produce tester scripts and DUT configurations, and assist in analyzing test results.
These capabilities pave the way for more automated, adaptive, and intelligent testing frameworks. In the following section, we discuss the key challenges in applying LLMs to automated network testing.

\subsection{Challenges}\label{subsec:challenges}

Despite the promising capabilities of LLMs, several key challenges remain in applying them to network protocol testing.

\myparagraph{C1: Understanding diverse and unstructured protocol specifications.}
Protocol specifications such as RFCs contain a wide variety of elements, ranging from abstract protocol semantics to low-level field and state machine definitions. 
These documents are lengthy, heterogeneous, and lack consistent formatting across different standards. 
As a result, LLMs struggle to fully capture both the high-level intent and the fine-grained technical details in one time, which are critical for generating valid test cases and device configurations.
To address this challenge, we combine high-level protocol function modeling with low-level detail modeling. 
This hybrid approach leverages the LLM's strength in language understanding while grounding it with structured protocol models, ensuring a more comprehensive and accurate interpretation of protocol specifications.

\myparagraph{C2: Evaluating the quality of test cases.}
Unlike source code, which has established metrics such as coverage or complexity, natural language test cases lack objective and standardized quality measures. 
Key aspects such as correctness, completeness, and coverage are difficult to quantify from free-form textual descriptions. 
Without proper evaluation metrics, automatically generated test cases may suffer from redundancy, gaps, or logical flaws, undermining their reliability.
We design a semi-quantitative evaluation mechanism that includes both section and semantic coverage analysis.

\myparagraph{C3: Translating natural language test cases into executable artifacts.}
Natural language descriptions cannot be directly executed on testing systems. 
They must be converted into multiple domain-specific representations, including scripts for testers and configuration files for the DUT. 
This translation process has traditionally relied heavily on expert knowledge and manual effort, creating a significant bottleneck in test automation.
To overcome this challenge, we design specialized LLM-based agents guided by expert knowledge and SOPs. These agents are responsible for generating domain-specific outputs in a modular fashion, ensuring correctness and adaptability across different testing platforms and device vendors.

\myparagraph{C4: Analyzing unified execution logs with heterogeneous error sources.}
A single execution log may simultaneously contain multiple types of error signals, such as DUT implementation bugs, DUT misconfigurations, mistakes in tester code generation, or flaws in test case design. 
The coexistence of these heterogeneous sources makes it difficult to directly identify root causes and provide meaningful feedback to earlier modules.
To address this challenge, we design a hierarchical feedback mechanism that enables step-by-step fault localization and offers targeted feedback to refine upstream modules efficiently.

\section{\methodName{} Overview}\label{sec:design-workflow}

\begin{figure*}[!t]
    \centering
    \includegraphics[width=0.91\textwidth]{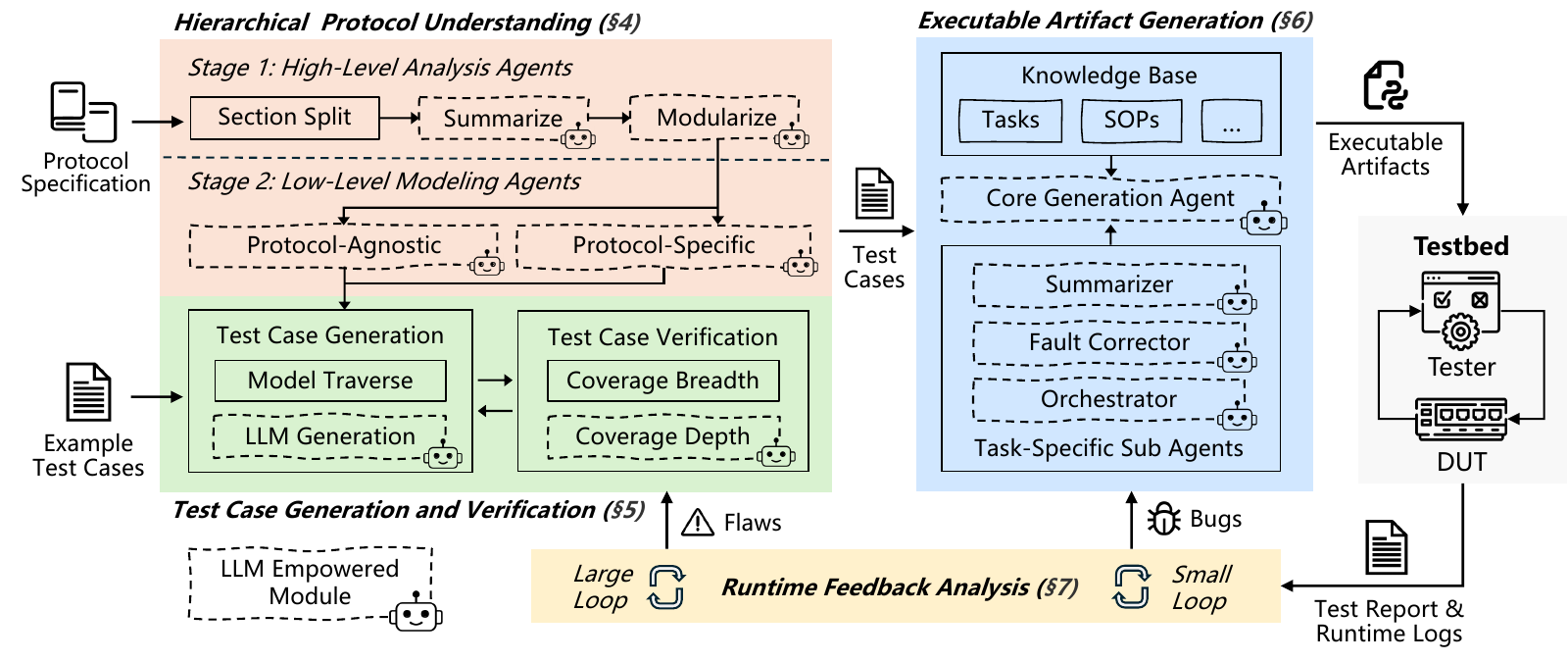}
    \vspace{-2ex}
    \caption{Workflow of \methodName{}. Modules marked with a robot icon are integrated with the LLM.}
    \label{fig:workflow}
\end{figure*}

The multi-agent workflow of \methodName{} is illustrated in Figure~\ref{fig:workflow}, which consists of four main components: hierarchical RFC understanding (\S~\ref{subsec:rfc-understanding}), test case generation with verification (\S~\ref{subsection:testcase-gen}), and executable artifact generation (\S~\ref{section:artifact-gen}), and runtime feedback analysis (\S~\ref{sec:runtime-feedback}).
For a target protocol specification, we first conduct hierarchical understanding and modeling.
The first stage is \textit{high-level analysis}, where the specification is divided into sections, each summarized and grouped into a set of protocol modules, with each module comprising its associated sections. 
The second stage is \textit{low-level modeling}, applied to each protocol module through protocol-agnostic or protocol-specific modeling.
Afterward, we traverse all module elements and, guided by high-quality reference test cases, employ LLMs to generate new test cases.
Finally, we evaluate the generated test cases in terms of coverage breadth (section coverage) and coverage depth (semantic coverage), and supplement missing cases as needed.

In the test case implement phase, an LLM agent generates the required artifacts, including tester scripts and DUT configuration files.
To enhance artifact generation, task-specific knowledge bases are built, comprising prompt templates, expert knowledge, and SOPs.
Beyond the general-purpose generation agent, we design several specialized sub-agents based on these expert knowledge tailored to downstream tasks, such as an experience-pool-driven fault-correction agent, a document and repository summarization agent, and an intent orchestration agent.
These resources collectively support the core LLM agent in generating high-quality artifacts.

The artifacts are executed in the testbed, whose feedback enables the agent to iteratively refine scripts and configurations for correctness (small loop).
Issues originating from the test cases themselves are further forwarded to the upstream test case generation phase for refinement (large loop).

\section{Hierarchical Protocol Understanding}\label{subsec:rfc-understanding}
Understanding protocol specifications is a prerequisite for generating test cases. 
Since network protocol standards (such as RFCs) are written for human readers rather than in machine-readable form, it is necessary to conduct hierarchical analysis and modeling to achieve comprehensive and in-depth understanding. 
To this end, we design a hierarchical pipeline for protocol understanding and modeling, consisting of two stages: high-level analysis and low-level modeling.

\subsection{High-Level Analysis Agents}\label{subsubsec:high-level-analysis}
Protocol specifications typically comprise multiple sections, each potentially covering different protocol functional modules.
The goal of high-level protocol analysis is to partition the complete specification document into distinct protocol functional modules, with each module linked to its corresponding section numbers.

\myparagraph{Section splitting.}
We construct a hierarchical RFC section tree by extracting metadata (RFC number, title, abstract, and table of contents) and section content, enabling a structured representation of the document (see Appendix~\ref{appendix:rfc_tree}).

\myparagraph{Section summarization.}
Given the length of protocol specifications, we adopt a section-wise summarization approach rather than processing the entire document at once, ensuring output quality and stability. 
We traverse the RFC section tree sequentially, using LLMs to generate concise summaries for each section, including references, classification (e.g., functional, descriptive, appendix, configuration), and test importance (high, medium, low). 
These outputs are stored in the corresponding RFC tree nodes. 
To improve contextual understanding, each prompt includes the RFC title, table of contents, and summaries of preceding sections, enabling the LLM to handle previously unseen protocols effectively.

\begin{figure}[!t]
    \centering
    \includegraphics[width=0.95\linewidth]{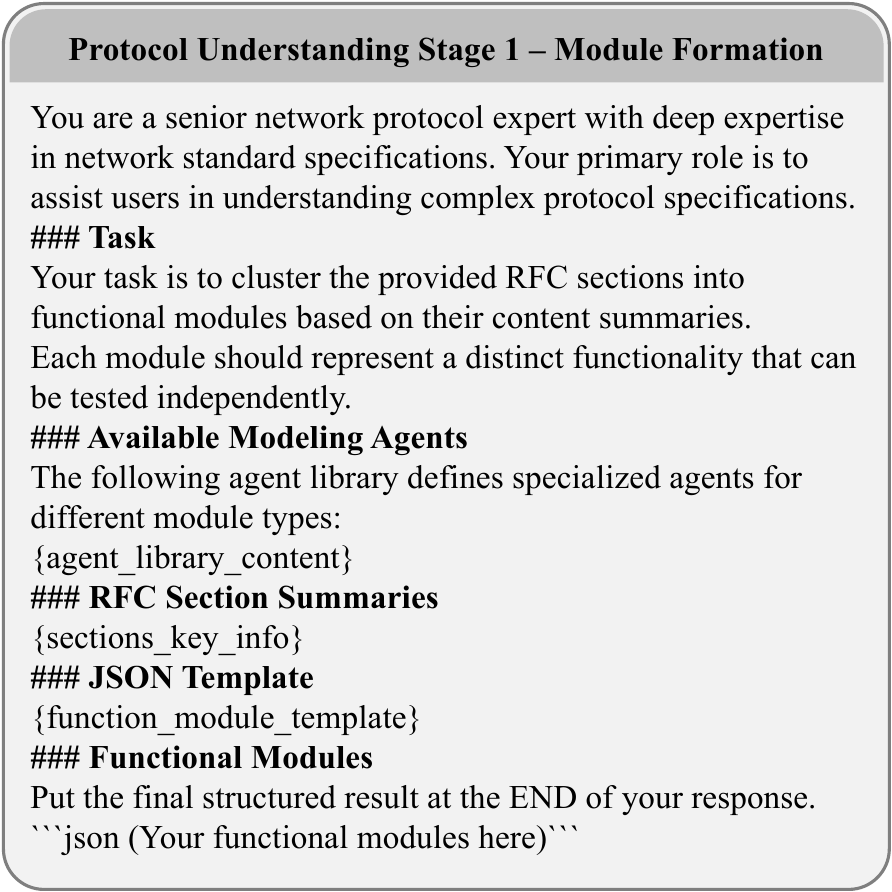}
    \vspace{-1ex}
    \caption{Prompt template for module formation.}
    \label{fig:prompt-module}
    \vspace{-1em}
\end{figure}

\myparagraph{Module formation.}
This is the core step of high-level analysis, aiming to partition sections into multiple protocol functional modules.
To align with subsequent low-level modeling, predefined agents are introduced, including: \textit{packet field modeling}, \textit{finite state machine (FSM) modeling}, \textit{time sequence modeling}, and \textit{protocol-specific function modeling} agent. 
Each agent is defined with its functionality, capabilities, and input/output specifications to enable accurate section-to-module mapping by the LLM.
The prompt for module formation includes: 
descriptions of the modeling agents, \textit{key information} of all sections of the protocol specification, and a JSON-based template specifying the output format. 
The \textit{key information} for each section includes the section number, title, summary, and testing importance score.
The prompt template is shown in Figure~\ref{fig:prompt-module}.

Since LLMs may overlook some important sections, we adopt an iterative completion mechanism.
After initial module formation, we use a rule-based method to identify uncovered sections.
We then use the \textit{key information} of all uncovered sections, along with the current module formation results as input, to prompt the LLM to generate a supplemented module formation result.
This process continues iteratively until no sections remain uncovered or a predefined maximum number of iterations is reached (set to 10 in practice).

\subsection{Low-Level Modeling Agents}
Low-level modeling constructs detailed representations of protocol modules to generate fine-grained, traversable test points for subsequent test case generation.
We categorize low-level modeling into two types: \textit{protocol-agnostic modeling} and \textit{protocol-specific modeling}. 
\textit{Protocol-agnostic modeling} refers to general modeling methods independent of a particular protocol, including packet-field modeling, FSM modeling, and message time sequence modeling. 
\textit{protocol-specific modeling} focuses on capturing unique functionalities of individual protocols, such as LSA flooding in OSPF or the decision process in BGP route selection.

\myparagraph{Protocol-agnostic structural modeling.}
Most protocols include common elements such as packet field, FSMs, and message time sequences. 
To capture such elements in a unified manner, we design three types of modeling agents: a \textit{packet field modeling} agent, an \textit{FSM modeling} agent, and a \textit{message time sequence} modeling agent.

The \textit{packet field modeling} agent transforms packet field definitions into structured data, capturing constraints and expected responses as testing points. 
Given a packet field module, the agent first reorders specification sections from header to body for logical consistency, and then incrementally traverses the reordered sections to extract structured field information until the module is fully covered.
The prompt template (see Appendix~\ref{appendix:low-level}) for extracting structured field information includes: task description, protocol summary, current and referenced section contents, previously extracted fields (if available), JSON template for field representation.

The \textit{FSM modeling} agent encodes FSM from protocol specifications into structured data. %
The FSM includes states and transitions.
Each transition includes source state, target state, triggering event, action, and constraints. %
The agent applies the following algorithm:

\vspace{-0.8em}
\begin{tcolorbox}[float=htbp,
  colback=white,     %
  colframe=black,    %
  boxrule=0.5pt,     %
  arc=3pt,           %
  boxsep=0pt,          %
  left=0pt, right=8pt, %
]
\begin{enumerate}[label=\circnum{\arabic*}, leftmargin=2em, itemsep=-1ex]
  \item Reviews all sections within the FSM module, extract as many states and transitions as possible, establishing a foundational framework for the FSM.
  \item Traverse the FSM module section by section.
  \item For each section, refine and supplement missing information based on local details.
  \item Incrementally integrate extracted transitions until all sections have been examined.
\end{enumerate}
\end{tcolorbox}
\vspace{-1em}

The \textit{message time sequence modeling} agent is designed to model message time sequences in the protocol as structured data, extracting information such as the order of message transmission and the expected responses. %
The design of this agent follows a similar methodology to that of packet-field modeling and FSM modeling. 
In contrast to FSM modeling, which focuses on state transitions within a single device, message-sequence modeling captures the temporal logic of message exchanges among multiple devices.

\begin{figure}[!t]
    \centering
    \includegraphics[width=0.87\linewidth]{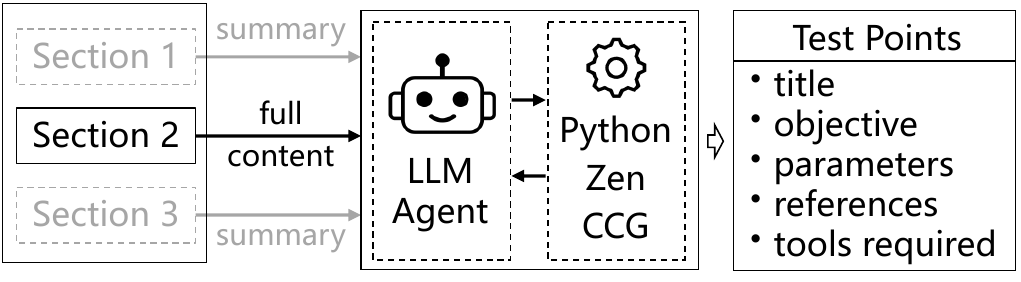}
    \vspace{-1.5ex}
    \caption{Workflow of protocol-specific functional modeling.}
    \label{fig:protocol-specific-modeling}
    \vspace{-1em} 
\end{figure}

\myparagraph{Protocol-specific functional modeling.}
Different protocols contain functional modules that are difficult to capture through structured modeling alone.
To address this, we design a general-purpose \textit{protocol-specific functional modeling} agent that flexibly extracts technical details and testing points from such modules.
As illustrated in Figure~\ref{fig:protocol-specific-modeling}, the agent employs a \textit{focus-moving mechanism} to balance detail and context: it sequentially traverses module sections, analyzing the full content of the current section while retaining summaries of others as contextual background.
The LLM then extracts candidate testing points from each section, represented as test case summaries with attributes such as a title, objective, parameters, and reference sections.
 
To enrich modeling depth and diversify test parameters, the agent also incorporates an extensible \textit{toolkit} of external resources. 
This toolkit includes programming languages (e.g., Python), the Zen constraint solver~\cite{beckett2020general, beckettZen} used in MESSI~\cite{singha2024messi}, and parsing tools such as Combinatory Categorial Grammar (CCG)\cite{artzi2013semantic} employed by SAGE\cite{yen2021semi}. 
Tool descriptions, including functionality and input/output specifications, are embedded in prompts to guide effective usage. 
In addition, the testing point template contains an \verb|additional_tools_required| field, enabling the LLM to specify suitable tools when necessary.

\section{Test Case Generation and Verification}
\label{subsection:testcase-gen}
\begin{figure}[!t]
    \centering
    \includegraphics[width=0.95\linewidth]{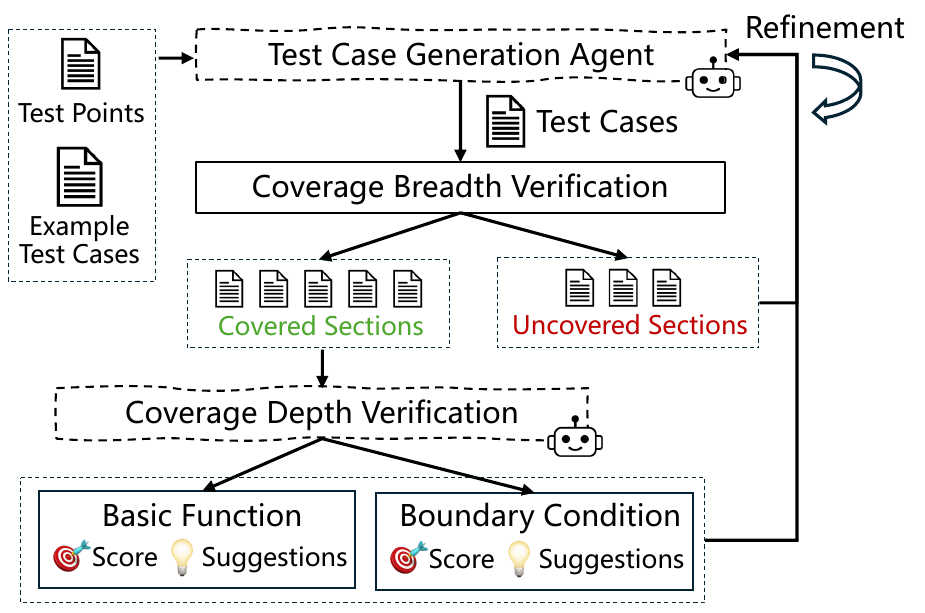}
    \vspace{-1.5ex}
    \caption{Iterative test case generation with verification.}
    \label{fig:tcg-ver}
    \vspace{-1em}
\end{figure}

Test case generation and verification is an iterative process, as illustrated in Figure~\ref{fig:tcg-ver}.
The LLM agent first generates test cases based on testing points derived from low-level modeling.
Next, the generated test cases are evaluated for coverage along two dimensions: breadth and depth.
Finally, the verification results are used to refine and supplement the test cases.

\subsection{Test Case Generation from Testing Points}

The objective of test case generation is to construct test cases that specify steps, expected results, and test topologies.     %
Given the success of LLMs in software test generation~\cite{wang2024hits, chen2024chatunitest, deng2024pentestgpt} and the language-oriented nature of network test cases, we adopt LLMs for this task.

The test case generation agent traverses all testing points extracted from low-level models (e.g., protocol fields, FSM transitions) and generates corresponding test cases in JSON format. 
Each case includes title, objective, steps, expected results, reference sections, and topology, with also a \verb|parameters| field for fine-grained implementation details. 
The prompt incorporates task description, test case template, current testing point, referenced content, and few-shot examples from industry or standard test cases to improve the quality and stylistic consistency of the generated test cases.
To further support new protocols, 
we also provide the protocol summary, metadata of the module to which the testing point belongs (including the module name and description), and summaries of relevant sections as context to the LLM.
The example prompt template is illustrated in Appendix~\ref{appendix:tcg_template}.

\subsection{Test Case Verification and Refinement}\label{subsubsec:testcase-verification}
Coverage and correctness are key metrics for evaluating generated test cases. Unlike software testing, where cases are formally specified in code, network protocol test cases are typically written in natural language, complicating the assessment of these metrics. In practice, we focus on improving coverage, while correctness is indirectly validated through subsequent code generation, execution, and refinement via the \textit{large loop} feedback mechanism.

\myparagraph{Coverage breadth verification.}
The breadth coverage of test cases is defined as the extent to which the generated test cases cover all important sections of the protocol specification.
Based on the testing importance and section classification from the high-level protocol analysis results (\S~\ref{subsubsec:high-level-analysis}), we design a threshold-based method to evaluate important sections.
The importance score of section $i$ is defined as:
{\setlength{\abovedisplayskip}{3pt}  %
\setlength{\belowdisplayskip}{3pt}   %
\begin{equation*}
score_i = test\_importance_i \times w(section\_classification_i)
\end{equation*}}
where $test\_importance$ and $section\_classification$ are derived from the high-level protocol analysis results, weight coefficient $w$ is a predefined mapping function that assigns weight values to different section categories.
The value of $test\_importance$ ranges from 0 to 100, and weight coefficient $w$ ranges from 0 to 1.
A threshold $\theta$ is then applied, such that all sections with $score_i \geq \theta$ are classified as important.
Both $weight\_coefficient$ and $\theta$ are tunable hyperparameters, 
which can be optimized to relax or tighten the selection criteria based on practical scenarios.
Finally, the coverage of all important sections by the generated test cases is calculated as the breadth metric for test case coverage.

\myparagraph{Coverage depth verification.}
We define coverage depth as the extent to which generated test cases address all test points within each important section of the protocol specification. 
To evaluate this, we adopt a dual-dimensional scoring framework: \textit{basic functionality coverage}, which measures how thoroughly essential features are tested, and \textit{boundary condition coverage}, which assesses parameter ranges, error handling, and extreme scenarios. 
We employ an LLM-as-a-judge~\cite{saha2025learning, gu2024survey, muhamed2025ccrszeroshotllmasajudgeframework} approach to assess coverage depth along these two dimensions.
The generated test cases are first grouped by section, after which the LLM evaluates each section's coverage.
Prompts include the task description, section content, associated test cases, scoring criteria, and a structured output template. 
The outputs provide both scores and rationales, along with concrete suggestions for improvement.

\myparagraph{Test case refinement.}
Based on the results of the two coverage evaluations, we further refine the generated test cases.
We first generate supplementary test cases for all uncovered important sections, followed by enhancing the depth of coverage for other sections.
This process can be iteratively repeated until the predefined coverage requirements are met or a maximum number of iterations is reached.

\section{Executable Artifact Generation}\label{section:artifact-gen}

To transform the generated test cases into executable artifacts, we designed a multi-round iterative generation system based on multi-agent collaboration and a domain-specific task knowledge base.
By centrally managing expert knowledge and domain expertise, and providing reusable sub-agents applicable to multiple tasks wherever possible, we aim to reduce the manual overhead of adapting to different domain-specific tasks, 
thereby assisting engineers in more efficiently converting test cases into executable artifacts.
The prompt templates guiding the artifact generation are detailed in Appendix~\ref{appendix:tsg_template}.

\subsection{Domain-Specific Task Knowledge Base}\label{subsec:task-knowledge-base}

The conversion of generated test cases into executable artifacts relies on multiple downstream tasks, including writing corresponding control scripts for the current test logic (e.g., test scripts executable on network testers) and generating configuration files for the DUT (e.g., command line interface (CLI) configuration for switches or routers). 
If the test environment lacks a simulation implementation for new protocols or protocol variations, protocol simulation also needs to be implemented.

These downstream tasks can be viewed as repository-level generation tasks but are highly domain-specific and typically depend on diverse expert knowledge. For example, generating test scripts based on testers requires testers to be familiar with the extensive testing APIs provided by the testers, while the configuration generation task for the DUT requires testers to be proficient in the complex CLI configuration methods of the device.
Therefore, we provide centralized management for different domain-specific tasks and integrate them into a domain-specific task knowledge base, which includes:
(1) \textit{task information}, including test case, task descriptions, repository structures, testbed devices, etc, (2) \textit{expert heuristics}, containing domain expert knowledge required by the LLM agents, and (3) \textit{SOPs}, containing implementation steps, code submission and review processes, etc.

\begin{figure}[t]
    \centering
    \includegraphics[width=0.87\linewidth]{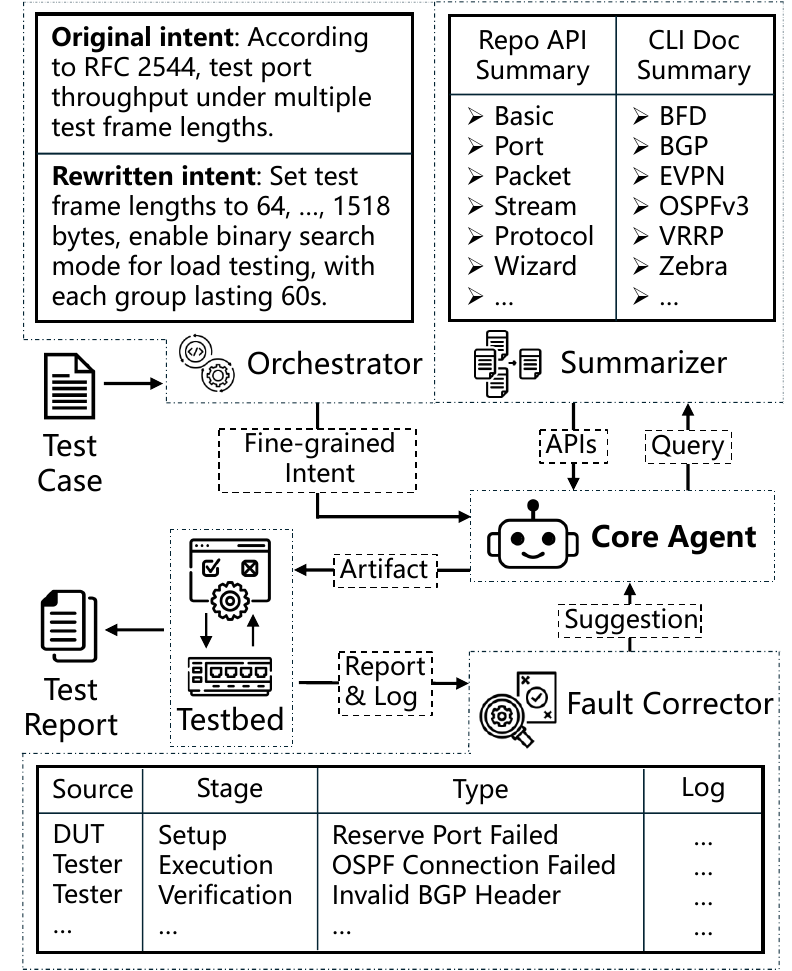}
    \caption{Workflow and examples of executable artifact generation agents.}
    \label{fig:tag-workflow}
\end{figure}

\subsection{Multi-Agent Artifact Generation}\label{subsec:multi-agent-artifact}

\methodName{} employs the mainstream generation agent~\cite{claude-code, gao2025trae, gemini-cli} as the core generation agent, with its workflow guided by SOPs. By enabling interaction with the test environment (e.g., submitting scripts for execution or deploying DUT configurations), the system allows for iterative refinement of generated results through multiple rounds to produce executable artifacts. To enhance domain adaptability, we have designed task-specific sub-agents that undergo continuous updates during the iterative process, thereby providing professional support for executable artifact generation. As shown in Figure~\ref{fig:tag-workflow}, these sub-agents include:

\myparagraph{Fault corrector.}
By recording the bugs encountered by the LLMs during generation and execution verification, along with their corresponding solutions, the LLM's performance can be enhanced when encountering similar errors in the future. Based on expert experience, various errors can be categorized to help the LLM better locate issues.

\myparagraph{Summarizer.}
Repository-level generation commonly relies on retrieval augmentation. For instance, the test script generation task for testers may require retrieving information from hundreds of testing APIs, while the configuration generation for the DUT depends on retrieving information from hundreds of pages of CLI documentation. Although most generation agents integrate retrieval capabilities, the LLM may still suffer from low retrieval efficiency due to a lack of expert knowledge. Therefore, following prior work~\cite{ma2024understand}, we generate summaries for entries in repositories or documents and construct a hierarchical index tree based on expert knowledge to better guide the LLM in retrieving the correct entries.

\myparagraph{Orchestrator.}
The test intent in textual test cases often includes test script logic, DUT configuration, and network topology at a coarse granularity.  
There is a need to extract fine-grained intents that can adapt to domain-specific tasks. To address this, for different domain-specific tasks, we use the LLM to extract corresponding fine-grained intents from existing test cases and their executable artifacts, serving as few-shot examples to guide the LLM in better performing task decomposition and intent extraction at the beginning.

Following prior work~\cite{wang2025towards}, we enable each sub-agent to learn from both successes and failures during iterative tasks and \textit{continuously update} its prompt templates or experience pools. 
The \textit{fault corrector} records issues that required multiple iterations to resolve into the experience pool when a test case is successfully generated, enabling faster fixes for similar errors in the future.
The \textit{summarizer} consults original repositories or documentation when it fails to retrieve relevant entries and updates the corresponding summary entries.
The \textit{orchestrator} adjusts its prompt templates based on final results, incorporating more effective few-shot examples or orchestration instructions.

\section{Runtime Feedback Analysis}\label{sec:runtime-feedback}

\myparagraph{Feedback for small loop.}
The small loop focuses on refining executable artifacts through iterative feedback.
In each round, the generated artifacts (e.g., tester scripts and CLI configuration files) are deployed to the testing environment and executed against the DUT.
The resulting runtime logs and reports are collected and returned to the executable artifact generation module.
A dedicated fault corrector sub-agent analyzes error messages and abnormal behaviors, classifies them according to common categories (e.g., syntax errors, configuration mismatches, unsupported commands), and retrieves similar historical cases to propose candidate fixes.
This loop enables rapid correction and re-deployment, ensuring that the artifacts evolve toward syntactically valid, semantically correct, and environment-compatible forms.
Through this process, the system minimizes redundant human intervention and progressively evolves into executable artifacts that faithfully embody the intent of the original test cases.

\myparagraph{Feedback for large loop.}
When repeated iterations in the small loop fail to produce executable artifacts that pass validation, the system escalates the issue to the large loop.
Unlike the small loop, which concentrates on artifact-level refinements, the large loop revisits the correctness of the underlying test cases and testing assumptions.
Specifically, unresolved errors may stem from multiple factors: (1) DUT implementation bugs or incomplete documentation, (2) functional limitations or defects in the tester, or (3) logical flaws or ambiguities in the test case design itself.
In this stage, the feedback is first routed back to the test case generation module, which may synthesize refined or alternative test cases to isolate the suspected cause.
If these regenerated cases still fail, the issue is flagged for manual review by human experts, who determine whether the failure originates from the DUT or from other systemic limitations.
This hierarchical escalation ensures that the framework not only corrects superficial artifact-level issues but also systematically addresses deeper inconsistencies in test design or device behavior.

\section{Implementation}\label{sec:implementation}

We implemented \methodName{} in Python with $\sim$8,000 lines of code (LoC) (without LLM prompts).  
For the protocol understanding and test case generation phase, we use OpenAI SDK~\cite{openai:openai-python} to access various LLMs and construct specialized agents.
In the executable artifact generation phase, we adopt Claude Code~\cite{claude-code} as the core general-purpose generation agent.

The overall design of \methodName{} is general and modular, enabling low-overhead adaptation to diverse standard documents (e.g., IETF and IEEE protocol standards or device specification documents), as well as flexible integration with different testing platforms (e.g., Ixia, Spirent, and Xinertel) and devices under test (e.g., Cisco, Juniper, and Huawei).  
Specifically, within the protocol understanding pipeline, only the \textit{section splitting} step in Stage~1 is document format dependent, requiring $\sim$140 LoC to implement, whereas the remaining components are fully generic.  
In Stage~2, the \textit{toolkit} adopts an extensible architecture, allowing new auxiliary modeling tools to be incorporated simply by adding their descriptions. Examples include programming languages, the Zen constraint-solving library~\cite{beckett2020general, beckettZen}, and the CCG parsing tool~\cite{artzi2013semantic}.  

In the executable artifact generation module, task-related expert knowledge, SOPs, and other content are recommended to be updated to better adapt to new downstream tasks. According to the experience from previous work~\cite{chen2022software}, switching equipment vendors typically requires an additional $\sim$50 LoC of scripting to crawl the device manuals and regenerate summaries. All other components and sub-agents can be reused.

\section{Evaluation}\label{sec:evaluation}

\subsection{Experimental Setup}\label{subsec:setup}

We use Qwen-Max~\cite{alibaba:qwen-max, qwen25} as the default LLM to build various agents in protocol understanding and test case generation phases.
Our evaluation covers three widely used protocols: OSPFv2 (RFC 2328), RIPv2 (RFC 2453), and BGP-4 (RFC 4271). 
We also utilized existing test cases from a specific power industry group as examples for our test case generation. 
The power industry has stringent requirements for test case quality due to its critical infrastructure nature, demanding high reliability and safety standards.

For executable artifact generation, we use GLM-4.5~\cite{zeng2025glm} as the default LLM of the agents.
To further assess the generation capabilities of different models, we also evaluated DeepSeek-V3.1~\cite{deepseekv31} and Qwen3-Coder-Plus~\cite{alibaba:qwen3-coder-plus}.
In the iterative generation process, the core agent need to verify that the produced scripts satisfy the expected execution results.
We set the maximum number of iteration rounds per attempt to 10, with up to 3 attempts.
If an artifact passes its own checks, it is then subject to manual review for final validation.

To cover different testing scenarios, we built two testbeds.  %
In one setup, we combined a Xinertel DARYU series network tester with a Huawei CE6881 switch (with routing capability) to emulate a typical acceptance testing environment.  
In the other, we paired a Xinertel BigTao series network tester with a host running FRRouting~\cite{frrouting:frr} (an open-source IP routing protocol suite), which better reflects a development-oriented environment where internal protocol behavior can be observed in detail.

\subsection{Dataset}

\myparagraph{FRRouting historical bug dataset.}  
To evaluate the effectiveness of our generated test cases on a real-world protocol implementation, we collected bug data from the GitHub repository of FRRouting. We crawled issues and historical commits and filtered those related to the three protocols we target in the test case generation stage (OSPFv2, RIPv2, and BGP-4). These bugs represent real defects, most of which have already been fixed. Such a dataset of real-world protocol implementation bugs provides strong evidence for assessing the coverage and depth achieved by different test case sets.

\myparagraph{National standard test suites.}  
To provide a baseline for evaluating our generated test cases, we employed existing national standard test suites, including YD/T 1251.2-2013~\cite{std:ospf-test} (OSPF protocol conformance testing methods), YD/T 1251.3-2013~\cite{std:bgp-test} (BGP protocol conformance testing methods), and YD/T 1156-2023~\cite{std:rip-test} (covering RIP protocol test cases).
These national standard test suites serve as the foundation for many enterprise test case collections and are widely used in the industry, making them highly valuable for comparison.

\myparagraph{Industrial test cases and executable artifacts.}
For executable artifact generation, we focused on two tasks: tester script generation and DUT configuration generation. 
we collected 38 industrial test cases together with their corresponding executable artifacts, including tester scripts and DUT configurations.
The tester scripts comprise a total of 1706 lines of Python code, which collectively invoke tester APIs 350 times during actual execution. The DUT configuration files contain a total of 1168 lines of CLI commands.
Among them, 10 cases were randomly selected for continuously updating the prompt templates and experience pools of the three sub-agents, while the remaining 29 were used as the test set.

\begin{figure}
  \centering
    \subfigure[Time cost.]{
        \includegraphics[width=0.145\textwidth]{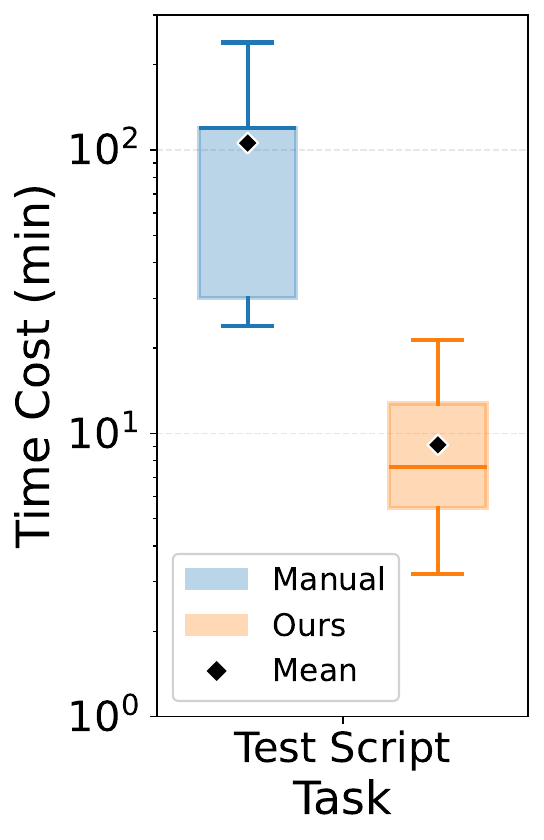}
        \label{fig:expert-tsg-time}
    }
    \subfigure[Expert evaluation results.]{
        \includegraphics[width=0.302\textwidth]{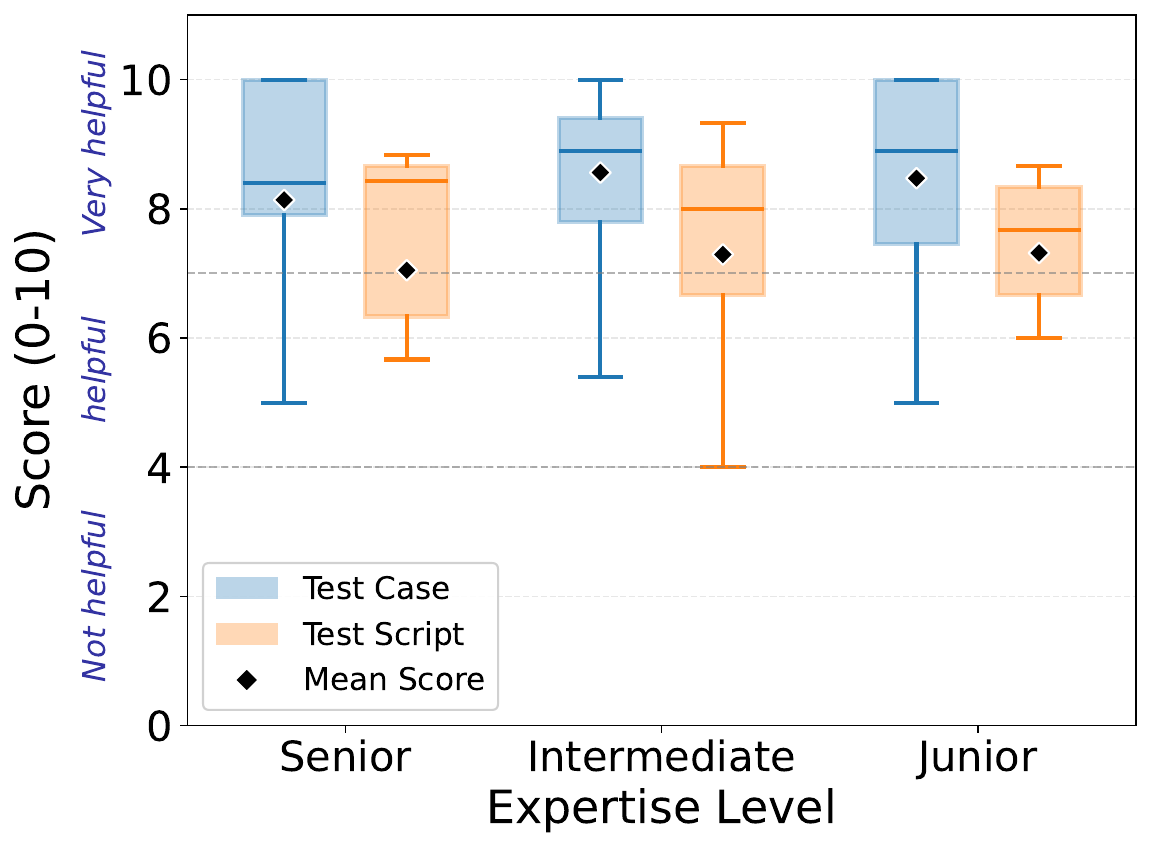}
        \label{fig:expert-tcg-tsg} 
    } 
    \vspace{-1ex}
  \caption{Time cost comparison and expert evaluation results.}
  \label{fig:expert-eval}
\end{figure}

\subsection{Production Result}\label{subsec:human}  %

We deployed \methodName{} in a production environment for several months. 
Through a survey of 12 testing domain experts who used \methodName{}, we compared it with manual methods across three dimensions: time, cost, and quality. This comparison demonstrates how \methodName{} saves time and economic costs while improving the quality of the generated test cases and executable artifacts. Subsequently, we present a case study to illustrate how \methodName{} operates in a typical scenario.
We used Qwen-Max~\cite{alibaba:qwen-max,qwen25} and GLM-4.5~\cite{zeng2025glm} as the default LLM for the test case generation and executable artifact generation stages, respectively.

\myparagraph{Time.}
The time costs of \methodName{} mainly depend on the response time of the LLM. The average time required for \methodName{} to generate one test case is 30.84 seconds. 
Although no statistical data are available, industry experience over many years suggests that manual test case writing typically takes hours. In comparison, \methodName{} 
achieves a speedup of approximately two orders of magnitude.
For test script generation, \methodName{} takes 9.10 minutes to generate one test script, while the average time from 12 experts is 1.74 hours (as shown in Figure~\ref{fig:expert-tsg-time}). %
The time required to manually refine the generated test scripts is approximately 12.07 minutes, thus enabling experts to achieve an 8.65x speedup (see \S~\ref{subsec:exec_arti_gene}). %

\myparagraph{Cost.}
The cost of \methodName{} also primarily stems from online inference for LLM agents. According to the latest token pricing of Qwen-Max~\cite{alibaba:qwen-max} and GLM-4.5~\cite{glm-4.5}, the cost for \methodName{} to generate one test case and one test script is \$0.0025 and \$0.81, respectively. In comparison, the average hourly wage for network test engineers in the United States in 2025 is \$52.42~\cite{network_test_engineer_hourly_wage}. Therefore, \methodName{} can reduce economic costs by several orders of magnitude.

\myparagraph{Quality.}
Following the survey methodology detailed in Appendix~\ref{appendix:survey}, we invited 12 domain experts to evaluate the generation quality of \methodName{}.
The evaluation results are presented in Figure~\ref{fig:expert-tcg-tsg}.
Across all expertise levels, our generated test cases consistently receive high scores, with both the median and average above 8 points , and an overall average of 8.40 (\textit{very helpful}). These results indicate that our generated test cases achieve high quality and are helpful to experts across different levels of expertise.
The generated scripts also receive positive feedback, with both the median and average above 7 points, and an overall average of 7.24 (\textit{very helpful}).
In addition, we ask the experts to directly compare generated scripts with the reference scripts, and 90\% of the samples are judged to have quality not lower than that of the reference scripts.

\myparagraph{Case study.}
A power industry purchaser requires performance and functional testing of its procured Layer 2 and Layer 3 switches. 
Traditionally, this process involves three parties (\S~\ref{subsec:motivation}): the power industry, as the testing service buyer, designs test cases according to its application requirements; the testing service provider translates these test cases into executable tester scripts; and the device vendor supplies corresponding configuration files for the DUTs. 
With \methodName{}, these tasks can be unified from the perspective of a \textit{neutral} testing service provider, which automatically generates test cases, tester scripts, and DUT configurations.   %
This not only improves testing efficiency but also  enhances fairness.

\begin{table*}[htbp]
  \centering
  \resizebox{0.9\textwidth}{!}{%
    \begin{tabular}{lcc|c|cccccc}  %
        \toprule
        \multirow{2}{*}{Protocol RFC} & \multicolumn{2}{c|}{\# of RFC Sections} & 
        \multirow{2}{*}{\shortstack{\# of Function \\Modules}} &  
        \multicolumn{6}{c}{\# of Generated Test Cases} \\
        & Total & Key for Test &  & Field & FSM & TimeSeq & Protocol-Specific & 1R-Supp. & Total \\
        \midrule
        OSPFv2 (RFC 2328) & 154 & 101 & 14 & 151 & 247 & 154 & 1295 & 896 & 2743 \\
        RIPv2 (RFC 2453)  & 36  & 20  & 7  & 27  & 190 & 95  & 154  & 174 & 640  \\
        BGP-4 (RFC 4271)  & 81  & 44  & 11 & 91  & 637 & 196 & 164  & 161 & 1249 \\
        \bottomrule
    \end{tabular}
  }
  \vspace{-1ex}
  \caption{Basic statistics of protocol test case generation process.}
  \label{tab:basic-gen-stats}
\end{table*}

\begin{table*}[htbp]
  \centering
  \resizebox{0.95\textwidth}{!}{%
    \begin{tabular}{lcc|cc|cc|cc}
        \toprule
        \multirow{2}{*}{Protocol RFC}  & 
        \multicolumn{2}{c|}{\makecell{Key Section \\ Coverage Rate}} & 
        \multicolumn{2}{c|}{\makecell{Basic Function \\ Coverage Score}} & 
        \multicolumn{2}{c|}{\makecell{Boundary Case \\ Coverage Score}} & 
        \multicolumn{2}{c}{\makecell{\# of FRRouting \\ Bug Covered}} \\
        & Industry Std. & \methodName{} & Industry Std. & \methodName{} & Industry Std. & \methodName{} & Industry Std. & \methodName{} \\
        \midrule
        OSPFv2 (RFC 2328) & 43.56\% & 99.01\% & 76.5 & 92.2 & 49.8 & 80.3 & 5 & 22 \\
        RIPv2 (RFC 2453)  & 60.00\% & 100.00\% & 89.0 & 91.4 & 65.3 & 81.3 & 0 & 4 \\
        BGP-4 (RFC 4271)  & 56.82\% & 95.45\% & 86.5 & 91.5 & 63.2 & 79.2 & 6 & 15 \\
        \bottomrule
    \end{tabular}
    }
    \vspace{-1ex}
  \caption{Comparison of test case coverage between industry standards and \methodName{}.}
  \label{tab:protocol-test-comparison}
  \vspace{-1ex}
\end{table*}

\subsection{Test Case Generation}

\myparagraph{Basic statistics of test case generation evaluation.}
We generate test cases for three protocols: OSPFv2, RIPv2, and BGP-4.  
The basic statistics of the test case generation process are summarized in Table~\ref{tab:basic-gen-stats}.  
The table reports data related to protocol understanding, including the total number of sections in each RFC, the number of key sections identified for testing, and the number of functional modules extracted from the RFCs (stage 1 of the hierarchical protocol understanding process).  
It also presents the number of test cases generated for different testing dimensions, including field testing, state machine testing, time sequence testing, and protocol-specific functional testing. 
Additionally, the table reports the number of test cases generated as a supplementary result following a one-round coverage verification (1R-Supp.). 
Finally, the total number of generated test cases is reported. 
\methodName{} generates hundreds to thousands of test cases for each protocol, covering various aspects of the protocols.

\myparagraph{RFC coverage.}
Following the test case evaluation method introduced in \S~\ref{subsubsec:testcase-verification}, we assess the coverage breadth and depth of the generated test cases and compare them with existing national standard test suites.
Coverage breadth is measured as the \textit{key section coverage rate}, indicating the proportion of key testing sections covered by the generated test cases. 
As shown in Table~\ref{tab:protocol-test-comparison}, our generated test cases substantially exceed the national standard test suites in terms of key section coverage.  
Coverage depth is evaluated using the \textit{basic functional coverage score} and the \textit{boundary case coverage score}. Although these two metrics are estimated by LLMs and do not provide absolute references, they serve as a basis for relative comparison among different test case sets and offer guidance for test case optimization. As reported in Table~\ref{tab:protocol-test-comparison}, our generated test cases achieve significantly higher scores than the national standard test suites in both basic functionality and boundary case coverage, indicating that our approach produces more comprehensive and in-depth test cases. 
On average, test cases generated by \methodName{} improve the basic functionality coverage score by 9.67\% and the boundary case coverage score by 37.02\% compared to the national standard.

\myparagraph{Ablation study.}
To validate that the test case verification and refinement process indeed enhances the quality of test cases, we conducted an ablation study.
Prior to verification and refinement, the boundary case coverage scores for OSPFv2, RIPv2, and BGP-4 were 73.8, 78.2, and 77.7, respectively. After one round of test case verification and refinement, these scores improved to 80.3, 81.3, and 79.2 (as shown in Table~\ref{tab:protocol-test-comparison}).
This single round of refinement resulted in an average improvement of 4.90\% in the boundary case coverage score, demonstrating that the supplementary test cases effectively enhance the coverage depth of the test suite.

\myparagraph{FRRouting bug coverage.}
To evaluate the error detection capability of \methodName{} in real-world protocol implementations, we conducted a coverage assessment using the constructed FRRouting historical bug dataset. 
The evaluation process involved keyword filtering, LLM pre-screening, and human review. 
The results are presented in Table~\ref{tab:protocol-test-comparison}.
The test cases generated by \methodName{} collectively cover 41 historical bugs across the three protocols, whereas the national standard test suites only cover 11 bugs. 
As a case study, we examine a bug in FRRouting (commit 4533dc6, Aug 21, 2020) where the BGP hold timer is incorrectly stopped in the OpenConfirm state, preventing the session from expiring and transitioning to Idle and violating RFC 4271.
Our generated test cases contains at least four cases that can reliably expose this defect.  
A typical example is a test case derived from the FSM transition on hold-timer expiry: the test sets up a session to OpenConfirm, withholds KEEPALIVE, and expects a NOTIFICATION (Hold Timer Expired) and transition to Idle. 
Due to the bug, the timer never fires and no notification is sent, causing the test to fail and thus revealing the implementation error.
These results demonstrate that \methodName{} possesses a strong capability for discovering bugs in protocol implementations.

\subsection{Executable Artifact Generation}\label{subsec:exec_arti_gene}

\myparagraph{Accuracy.}
As described in \S~\ref{subsec:setup}, a manual review is conducted on the generated executable artifacts to calculate the \textit{Validation Rate} (VR). \textit{Recall} (R) computes the line-by-line recall rate between the generated results and the standard answers, i.e., how many lines in the standard answers are included in the generated results. For script generation, we define a "line" as one API call in the execution results to evaluate the behavioral consistency of the script. For configuration generation, a "line" refers to one CLI command, ignoring empty lines, comments, or common equivalent expressions (e.g., \texttt{ip address 10.0.0.1/24} and \texttt{ip address 10.0.0.1 255.255.255.0} are equivalent). To assess how many lines network engineers need to add, delete, or replace in the generated results, the \textit{Similarity Score} (SIM) is calculated based on the \textit{Normalized Edit Distance} (NED). The formulas for the above two metrics are as follows:
{\setlength{\abovedisplayskip}{7pt}  %
\setlength{\belowdisplayskip}{7pt}   %
\begin{equation*}
SIM=1-NED=1-\frac{\text{edit distance}}{\max(len_{ans},len_{out})}
\end{equation*}
}

The results are shown in Table~\ref{tab:exec-arti-gene}. \methodName{} achieves validation rates of 89.7\% and 93.1\% for the script generation and configuration generation tasks, respectively. The recall rates reach 91.6\% and 90.6\%, indicating high consistency between the generated results and the standard answers. SIM scores of 72.4\% and 76.9\% suggest that network engineers need to modify approximately 27.6\% and 23.1\% of the content for the two tasks, respectively.%

\begin{table}[t]
  \centering 
  \resizebox{0.47\textwidth}{!}{%
    \begin{tabular}{lcccc|cc}
        \toprule
        \multirow{2}{*}{Task} & \multirow{2}{*}{Sub-agents}
        & \multicolumn{3}{c|}{Accuracy} & \multicolumn{2}{c}{Time\ (min)} \\
        && VR & R & SIM & Infer. & Total \\
        \midrule
        \multirow{5}{*}{Script}     & F,S,O & 89.7\% & 91.6\% & 72.4\% & 5.24 & 9.10 \\
                                    & F     & 86.2\% & 89.8\% & 70.1\% & 5.45 & 10.38 \\
                                    & S     & 72.4\% & 83.6\% & 59.4\% & 5.37 & 10.21 \\
                                    & O     & 69.0\% & 83.0\% & 59.2\% & 6.78 & 11.28 \\
                                    & /     & 65.5\% & 75.9\% & 46.6\% & 9.00 & 13.27 \\
        \midrule
        \multirow{5}{*}{Config}     & F,S,O & 93.1\% & 90.6\% & 76.9\% & 3.03 & 5.62 \\
                                    & F     & 89.7\% & 87.3\% & 73.5\% & 3.29 & 5.81 \\
                                    & S     & 79.3\% & 89.7\% & 62.3\% & 3.07 & 6.43 \\
                                    & O     & 69.0\% & 81.5\% & 54.9\% & 4.02 & 6.84 \\
                                    & /     & 69.0\% & 82.2\% & 55.1\% & 4.33 & 7.08 \\
        \bottomrule
        \multicolumn{7}{c}{\small F,S,O represent fault corrector, summarizer and orchestrator respectively.}  %
    \end{tabular}
    }
    \vspace{-0.5ex}
  \caption{Performance of executable artifact generation.}
  \label{tab:exec-arti-gene}
\end{table}

\myparagraph{Time cost.}
To evaluate its practical acceleration effect, we assess the average generation time per result (\textit{Total}) with the LLM inference time (\textit{Infer.}).
The results are shown in Table~\ref{tab:exec-arti-gene}. The average completion times for the two tasks are 9.10 and 5.62 minutes respectively, with LLM inference times of 5.24 and 3.03 minutes respectively. In comparison, the estimated time required for manually writing scripts is 1.74 hours (see \S~\ref{subsec:human}). Consequently, the expected time for modifying results generated by \methodName{} is {$9.10\text{min}+(1-VR)\times (1-SIM) \times 1.74\text{h}=12.07$} minutes, thereby assisting network engineers in achieving an acceleration ratio of approximately 8.65x.

\myparagraph{Ablation study.}
We conducted an ablation study on the three sub-agents, with the results shown in Table~\ref{tab:exec-arti-gene}. Incorporating all three sub-agents improved the validation rates for the two tasks from 65.5\% and 69.0\% to 89.7\% and 93.1\%, representing improvements of 24.2\%. The average time was reduced from 13.27 and 7.08 minutes to 9.10 and 5.62 minutes, achieving acceleration ratios of 1.46x and 1.26x. All three sub-agents contribute to varying degrees of performance improvement, with the fault corrector and summarizer demonstrating the most significant impact on performance enhancement.

\myparagraph{Running with other LLMs.}
To validate the versatility of \methodName{}, we also conducted tests on DeepSeek-V3.1~\cite{deepseekv31} and Qwen3-Coder-Plus~\cite{alibaba:qwen3-coder-plus}. The results, as shown in Table~\ref{tab:diff-models}, demonstrate stable performance across other mainstream models, indicating that the generation system of \methodName{} is not reliant on any specific LLM.

\begin{table}[t]
    \centering
    \resizebox{0.77\columnwidth}{!}{%
        \begin{tabular}{llcc}
        \toprule
        Task & Model Name & VR & Time\ (min) \\
        \midrule
        \multirow{3}{*}{Script} & GLM-4.5           & 89.7\% & 9.10 \\
                                & DeepSeek-V3.1     & 79.3\% & 8.49 \\
                                & Qwen3-Coder-Plus  & 86.2\% & 10.41 \\
        \midrule
        \multirow{3}{*}{Config} & GLM-4.5           & 93.1\% & 5.62 \\
                                & DeepSeek-V3.1     & 86.2\% & 5.54 \\
                                & Qwen3-Coder-Plus  & 93.1\% & 6.67 \\
        \bottomrule
        \end{tabular}
    }
    \vspace{-0.5ex}

    \caption{Executable artifact generation with different LLMs.}
    \label{tab:diff-models} 
\end{table}

\section{Limitation and Future Work}\label{sec:limitation}

\myparagraph{Hybrid test case evaluation mechanism.}
\methodName{} currently leverages LLM-as-a-judge to iteratively enhance test coverage. Although this approach does not rely on formal theoretical models, it provides insightful suggestions that effectively guide test case refinement. 
More broadly, test case evaluation can be viewed as a meaningful and independent research direction, whose advances can benefit both automated generation and human-written test cases. 
Future work can explore integrating protocol models and formal methods to establish a more systematic evaluation framework, further supporting high-quality test case generation.

\myparagraph{Adaptive test case generation from failure analysis.}
At present, \methodName{} aims to refine test cases iteratively using feedback to enhance correctness. 
This represents a key step toward adaptive testing. %
The broader objective of automated testing is to adaptively discover vulnerabilities and defects in protocol implementations.
A promising research direction is to dynamically and selectively generate deeper test cases based on failure reports, thereby facilitating more effective and targeted vulnerability discovery.

\myparagraph{Test case suites management and optimization.}
While \methodName{} already generates a broad and diverse set of test cases with strong practical applicability, further efficiency gains can be achieved through test case suite management and optimization.
For example, the testing process involves setting up the test environment according to the requirements of each test case, which primarily consists of topology configuration. 
Identifying a minimal common topology that can satisfy multiple test cases could significantly improve test environment construction efficiency.
This represents a promising future direction for enhancing the automated testing system.

\myparagraph{Enhancing artifact generation through model specialization.}
The current \methodName{} generates executable artifacts through a multi-agent approach without LLM training or fine-tuning. 
While incorporating reinforcement learning with human feedback (RLHF)\cite{ouyang2022training} or the latest reinforcement learning with verifiable rewards (RLVR)\cite{lambert2024tulu} could enable the development of industry-standard domain-specific models, such approaches may incur higher time, economic, and human resource costs, particularly when switching equipment suppliers. 
Future work includes integrating these approaches into \methodName{} while exploring how to balance task-specific performance with these costs.

\section{Related Work}\label{sec:related}
\label{sec:related-work}

\myparagraph{Network protocol understanding.}
Some existing explored automated approaches to understand protocol specifications.
SAGE~\cite{yen2021semi} employs a semantic parsing technique (CCG~\cite{artzi2013semantic}) to disambiguate protocol specifications and generate code,
but it is a rule-based method that lacks scalability.
Recent studies have explored using NLP techniques to understand protocol specifications.
RFCNLP~\cite{pacheco2022automated} uses a data-driven hybrid approach to extract FSMs from RFC specifications for protocol security.
PROSPER~\cite{sharma2023prosper} leverages LLMs to extract protocol specifications from RFC documents.

\myparagraph{Test case generation.}
Some model-based methods have been proposed for generating test cases.
For example,
SCALE~\cite{kakarla2022scale} applies symbolic execution of an executable DNS model to detect RFC compliance errors, and MESSI~\cite{singha2024messi} extends it with modular exploration to handle stateful protocols such as BGP.
However, these methods lack protocol generalization ability.
More recently, LLM-based methods have been introduced for security testing.
For example, ChatAFL~\cite{meng2024large} leverages LLMs to augment existing mutation-based protocol fuzzing, and PenTestGPT~\cite{deng2024pentestgpt} automates end-to-end penetration testing with a multi-module LLM architecture.

\myparagraph{Network configuration generation.}
Recent studies have explored using LLMs for network configuration generation.
Verified Prompt Programming~\cite{mondal2023llms} integrates GPT-4 with verifiers to iteratively generate correct router configurations. 
CEGS~\cite{liu2025cegs} automates network configuration synthesis with graph neural networks (GNNs) and LLMs based on configuration examples. %
Confucius~\cite{wang2025intent} uses a multi-agent LLM framework to generate network configuration.

\myparagraph{Repository-level code generation.}
Retrieval-augmented code generation (RaCG) methods~\cite{zhou2022docprompting, ma2024compositional, zan2023private, li2024repomincoder, liao20243} generate repository-level code by retrieving and extracting repository information.
For example, A$^3$-CodGen~\cite{liao20243} integrates local, global, and third-party libraries to improve code reusability.
However, most RaCG methods rely on single-round generation and underutilize interactive development environments.
To address this, recent work has attempted to use LLM agent techniques: SWE-agent~\cite{yang2024swe} studies the impact of terminal interface design, Trae Agent~\cite{gao2025trae} introduces the first agent-based ensemble reasoning approach for repository-level issue resolution, and several projects~\cite{claude-code, gemini-cli, qwen-code, trae-agent} have facilitated practical adoption.  %

This paper is an extended version of~\cite{10.1145/3744200.3744763}. 
Compared to this workshop paper, we enhanced the architecture with a hierarchical protocol understanding pipeline and multi-agent artifact generation system, added semi-quantitative test case verification, and hierarchical runtime feedback. 
We also expanded evaluations through FRRouting bug coverage, comparison with national standards, and expert evaluation, 
demonstrating the efficiency and practicality of the framework.

\section{Conclusion}\label{sec:conclusion}
This paper presented \methodName{}, a multi-agent LLM-based framework for automated network protocol testing that integrates specification comprehension, test case generation, artifact translation, and runtime log analysis. 
Experimental results on mainstream routing protocols and historical bug datasets demonstrate that \methodName{} improves testing coverage. %
Expert evaluations further show that the framework reduces human effort and produces high-quality test cases and executable artifacts.
These results highlight \methodName{}’s capability as a scalable solution for adapting protocol testing to evolving standards and heterogeneous devices.

\bibliographystyle{plain}
\bibliography{references}

\begin{thebibliography}{10}

\bibitem{std:bgp-test}
Border gateway protocol (bgp4): A conformance testing method for routing protocols.
\newblock \url{https://std.samr.gov.cn/hb/search/stdHBDetailed?id=8B1827F1BED1BB19E05397BE0A0AB44A}, 2013.

\bibitem{std:ospf-test}
Open shortest path first protocol (ospf): A conformance testing method for routing protocols.
\newblock \url{https://std.samr.gov.cn/hb/search/stdHBDetailed?id=8B1827F1B943BB19E05397BE0A0AB44A}, 2013.

\bibitem{std:rip-test}
Router device test method: Core router.
\newblock \url{https://std.samr.gov.cn/hb/search/stdHBDetailed?id=108B29E379F4B367E06397BE0A0AAFC2}, 2023.

\bibitem{network_test_engineer_hourly_wage}
Average hourly wage of network test engineer in the united states of 2025.
\newblock \url{https://www.ziprecruiter.com/Salaries/Network-Test-Engineer-Salary#Hourly}, 2025.

\bibitem{alibaba:qwen-max}
{Alibaba Group.}
\newblock {Qwen-Max.}
\newblock \url{https://bailian.console.aliyun.com/model-market/detail/qwen-max#/model-market/detail/qwen-max}, 2025.

\bibitem{alibaba:qwen3-coder-plus}
{Alibaba Group.}
\newblock {Qwen3-Coder-Plus.}
\newblock \url{https://bailian.console.aliyun.com/?tab=model#/model-market/detail/group-qwen3-coder-plus}, 2025.

\bibitem{claude-code}
{Anthropic.}
\newblock Claude code.
\newblock \url{https://www.anthropic.com/claude-code}.

\bibitem{artzi2013semantic}
Yoav Artzi, Nicholas FitzGerald, and Luke Zettlemoyer.
\newblock Semantic parsing with combinatory categorial grammars.
\newblock {\em ACL (Tutorial Abstracts)}, 3, 2013.

\bibitem{beckett2020general}
Ryan Beckett and Ratul Mahajan.
\newblock A general framework for compositional network modeling.
\newblock In {\em Proceedings of the 19th ACM Workshop on Hot Topics in Networks}, pages 8--15, 2020.

\bibitem{glm-4.5}
{BigModel.}
\newblock {GLM-4.5.}
\newblock \url{https://bigmodel.cn/console/modelcenter/modeldetails/10236}, 2025.

\bibitem{trae-agent}
{bytedance.}
\newblock Trae agent.
\newblock \url{https://github.com/bytedance/trae-agent}.

\bibitem{chen2022software}
Huangxun Chen, Yukai Miao, Li~Chen, Haifeng Sun, Hong Xu, Libin Liu, Gong Zhang, and Wei Wang.
\newblock Software-defined network assimilation: bridging the last mile towards centralized network configuration management with nassim.
\newblock In {\em Proceedings of the ACM SIGCOMM 2022 Conference}, pages 281--297, 2022.

\bibitem{chen2024chatunitest}
Yinghao Chen, Zehao Hu, Chen Zhi, Junxiao Han, Shuiguang Deng, and Jianwei Yin.
\newblock Chatunitest: A framework for llm-based test generation.
\newblock In {\em Companion Proceedings of the 32nd ACM International Conference on the Foundations of Software Engineering}, pages 572--576, 2024.

\bibitem{deepseekv31}
{DeepSeek.}
\newblock {DeepSeek-V3.1 Release.}
\newblock \url{https://api-docs.deepseek.com/news/news250821}, 2025.

\bibitem{deng2024pentestgpt}
Gelei Deng, Yi~Liu, V{\'\i}ctor Mayoral-Vilches, Peng Liu, Yuekang Li, Yuan Xu, Tianwei Zhang, Yang Liu, Martin Pinzger, and Stefan Rass.
\newblock $\{$PentestGPT$\}$: Evaluating and harnessing large language models for automated penetration testing.
\newblock In {\em 33rd USENIX Security Symposium (USENIX Security 24)}, pages 847--864, 2024.

\bibitem{frrouting:frr}
{FRRouting.}
\newblock {frr.}
\newblock \url{https://github.com/FRRouting/frr}, 2025.

\bibitem{gao2025trae}
Pengfei Gao, Zhao Tian, Xiangxin Meng, Xinchen Wang, Ruida Hu, Yuanan Xiao, Yizhou Liu, Zhao Zhang, Junjie Chen, Cuiyun Gao, et~al.
\newblock Trae agent: An llm-based agent for software engineering with test-time scaling.
\newblock {\em arXiv preprint arXiv:2507.23370}, 2025.

\bibitem{gemini-cli}
{google-gemini.}
\newblock Gemini cli.
\newblock \url{https://github.com/google-gemini/gemini-cli}.

\bibitem{gu2024survey}
Jiawei Gu, Xuhui Jiang, Zhichao Shi, Hexiang Tan, Xuehao Zhai, Chengjin Xu, Wei Li, Yinghan Shen, Shengjie Ma, Honghao Liu, et~al.
\newblock A survey on llm-as-a-judge.
\newblock {\em arXiv preprint arXiv:2411.15594}, 2024.

\bibitem{hong2024metagpt}
Sirui Hong, Mingchen Zhuge, Jonathan Chen, Xiawu Zheng, Yuheng Cheng, Ceyao Zhang, Jinlin Wang, Zili Wang, Steven Ka~Shing Yau, Zijuan Lin, et~al.
\newblock Metagpt: Meta programming for a multi-agent collaborative framework.
\newblock International Conference on Learning Representations, ICLR, 2024.

\bibitem{kakarla2022scale}
Siva Kesava~Reddy Kakarla, Ryan Beckett, Todd Millstein, and George Varghese.
\newblock $\{$SCALE$\}$: Automatically finding $\{$RFC$\}$ compliance bugs in $\{$DNS$\}$ nameservers.
\newblock In {\em 19th USENIX Symposium on Networked Systems Design and Implementation (NSDI 22)}, pages 307--323, 2022.

\bibitem{lai2025leocc}
Zeqi Lai, Zonglun Li, Qian Wu, Hewu Li, Jihao Li, Xin Xie, Yuanjie Li, Jun Liu, and Jianping Wu.
\newblock Leocc: Making internet congestion control robust to leo satellite dynamics.
\newblock In {\em Proceedings of the ACM SIGCOMM 2025 Conference}, pages 129--146, 2025.

\bibitem{lambert2024tulu}
Nathan Lambert, Jacob Morrison, Valentina Pyatkin, Shengyi Huang, Hamish Ivison, Faeze Brahman, Lester James~V Miranda, Alisa Liu, Nouha Dziri, Shane Lyu, et~al.
\newblock Tulu 3: Pushing frontiers in open language model post-training.
\newblock {\em arXiv preprint arXiv:2411.15124}, 2024.

\bibitem{li2024repomincoder}
Yifan Li, Ensheng Shi, Dewu Zheng, Kefeng Duan, Jiachi Chen, and Yanlin Wang.
\newblock Repomincoder: Improving repository-level code generation based on information loss screening.
\newblock In {\em Proceedings of the 15th Asia-Pacific Symposium on Internetware}, pages 229--238, 2024.

\bibitem{liao20243}
Dianshu Liao, Shidong Pan, Xiaoyu Sun, Xiaoxue Ren, Qing Huang, Zhenchang Xing, Huan Jin, and Qinying Li.
\newblock A 3-codgen: A repository-level code generation framework for code reuse with local-aware, global-aware, and third-party-library-aware.
\newblock {\em IEEE Transactions on Software Engineering}, 2024.

\bibitem{liu2025cegs}
Jianmin Liu, Li~Chen, Dan Li, and Yukai Miao.
\newblock {CEGS}: Configuration example generalizing synthesizer.
\newblock In {\em 22nd USENIX Symposium on Networked Systems Design and Implementation (NSDI 25)}, pages 1327--1347, 2025.

\bibitem{liu2024democratizing}
Lixin Liu, Yuanjie Li, Hewu Li, Jiabo Yang, Wei Liu, Jingyi Lan, Yufeng Wang, Jiarui Li, Jianping Wu, Qian Wu, et~al.
\newblock Democratizing direct-to-cell low earth orbit satellite networks.
\newblock {\em GetMobile: Mobile Computing and Communications}, 28(2):5--10, 2024.

\bibitem{ma2024understand}
Yingwei Ma, Qingping Yang, Rongyu Cao, Binhua Li, Fei Huang, and Yongbin Li.
\newblock How to understand whole software repository?
\newblock {\em arXiv preprint arXiv:2406.01422}, 2024.

\bibitem{ma2024compositional}
Zexiong Ma, Shengnan An, Bing Xie, and Zeqi Lin.
\newblock Compositional api recommendation for library-oriented code generation.
\newblock In {\em Proceedings of the 32nd IEEE/ACM International Conference on Program Comprehension}, pages 87--98, 2024.

\bibitem{meng2024large}
Ruijie Meng, Martin Mirchev, Marcel B{\"o}hme, and Abhik Roychoudhury.
\newblock Large language model guided protocol fuzzing.
\newblock In {\em Proceedings of the 31st Annual Network and Distributed System Security Symposium (NDSS)}, volume 2024, 2024.

\bibitem{mondal2023llms}
Rajdeep Mondal, Alan Tang, Ryan Beckett, Todd Millstein, and George Varghese.
\newblock What do llms need to synthesize correct router configurations?
\newblock In {\em Proceedings of the 22nd ACM Workshop on Hot Topics in Networks}, pages 189--195, 2023.

\bibitem{muhamed2025ccrszeroshotllmasajudgeframework}
Aashiq Muhamed.
\newblock {CCRS}: A zero-shot llm-as-a-judge framework for comprehensive rag evaluation, 2025.

\bibitem{openai:openai-python}
{openai.}
\newblock {openai-python.}
\newblock \url{https://github.com/openai/openai-python}, 2025.

\bibitem{ouyang2022training}
Long Ouyang, Jeffrey Wu, Xu~Jiang, Diogo Almeida, Carroll Wainwright, Pamela Mishkin, Chong Zhang, Sandhini Agarwal, Katarina Slama, Alex Ray, et~al.
\newblock Training language models to follow instructions with human feedback.
\newblock {\em Advances in neural information processing systems}, 35:27730--27744, 2022.

\bibitem{pacheco2022automated}
Maria~Leonor Pacheco, Max von Hippel, Ben Weintraub, Dan Goldwasser, and Cristina Nita-Rotaru.
\newblock Automated attack synthesis by extracting finite state machines from protocol specification documents.
\newblock In {\em 2022 IEEE Symposium on Security and Privacy (SP)}, pages 51--68. IEEE, 2022.

\bibitem{qwen-code}
{QwenLM.}
\newblock Qwen code.
\newblock \url{https://github.com/QwenLM/qwen-code}.

\bibitem{beckettZen}
{Ryan Beckett.}
\newblock Zen.
\newblock \url{https://github.com/microsoft/Zen/tree/master}.

\bibitem{saha2025learning}
Swarnadeep Saha, Xian Li, Marjan Ghazvininejad, Jason~E Weston, and Tianlu Wang.
\newblock Learning to plan \& reason for evaluation with thinking-{LLM}-as-a-judge.
\newblock In {\em Forty-second International Conference on Machine Learning}, 2025.

\bibitem{sharma2023prosper}
Prakhar Sharma and Vinod Yegneswaran.
\newblock Prosper: Extracting protocol specifications using large language models.
\newblock In {\em Proceedings of the 22nd ACM Workshop on Hot Topics in Networks}, pages 41--47, 2023.

\bibitem{singha2024messi}
Rathin Singha, Rajdeep Mondal, Ryan Beckett, Siva Kesava~Reddy Kakarla, Todd Millstein, and George Varghese.
\newblock $\{$MESSI$\}$: Behavioral testing of $\{$BGP$\}$ implementations.
\newblock In {\em 21st USENIX Symposium on Networked Systems Design and Implementation (NSDI 24)}, pages 1009--1023, 2024.

\bibitem{qwen25}
Qwen Team.
\newblock Qwen2.5 technical report.
\newblock {\em arXiv preprint arXiv:2412.15115}, 2024.

\bibitem{wang2025towards}
Chenxu Wang, Xumiao Zhang, Runwei Lu, Xianshang Lin, Xuan Zeng, Xinlei Zhang, Zhe An, Gongwei Wu, Jiaqi Gao, Chen Tian, et~al.
\newblock Towards llm-based failure localization in production-scale networks.
\newblock In {\em Proceedings of the ACM SIGCOMM 2025 Conference}, pages 496--511, 2025.

\bibitem{wang2024hits}
Zejun Wang, Kaibo Liu, Ge~Li, and Zhi Jin.
\newblock Hits: High-coverage llm-based unit test generation via method slicing.
\newblock In {\em Proceedings of the 39th IEEE/ACM International Conference on Automated Software Engineering}, pages 1258--1268, 2024.

\bibitem{wang2025intent}
Zhaodong Wang, Samuel Lin, Guanqing Yan, Soudeh Ghorbani, Minlan Yu, Jiawei Zhou, Nathan Hu, Lopa Baruah, Sam Peters, Srikanth Kamath, et~al.
\newblock Intent-driven network management with multi-agent llms: The confucius framework.
\newblock In {\em Proceedings of the ACM SIGCOMM 2025 Conference}, pages 347--362, 2025.

\bibitem{10.1145/3744200.3744763}
Yunze Wei, Kaiwen Chi, Shibo Du, Xiaohui Xie, Ziyu Geng, Yuwei Han, Zhen Li, Zhanyou Li, and Yong Cui.
\newblock Large language model driven automated network protocol testing.
\newblock In {\em Proceedings of the 2025 Applied Networking Research Workshop}, ANRW '25, page 32–38, New York, NY, USA, 2025. Association for Computing Machinery.

\bibitem{wu2024autogen}
Qingyun Wu, Gagan Bansal, Jieyu Zhang, Yiran Wu, Beibin Li, Erkang Zhu, Li~Jiang, Xiaoyun Zhang, Shaokun Zhang, Jiale Liu, et~al.
\newblock Autogen: Enabling next-gen llm applications via multi-agent conversations.
\newblock In {\em First Conference on Language Modeling}.

\bibitem{yang2024swe}
John Yang, Carlos Jimenez, Alexander Wettig, Kilian Lieret, Shunyu Yao, Karthik Narasimhan, and Ofir Press.
\newblock Swe-agent: Agent-computer interfaces enable automated software engineering.
\newblock {\em Advances in Neural Information Processing Systems}, 37:50528--50652, 2024.

\bibitem{yen2021semi}
Jane Yen, Tam{\'a}s L{\'e}vai, Qinyuan Ye, Xiang Ren, Ramesh Govindan, and Barath Raghavan.
\newblock Semi-automated protocol disambiguation and code generation.
\newblock In {\em Proceedings of the 2021 ACM SIGCOMM 2021 Conference}, pages 272--286, 2021.

\bibitem{zan2023private}
Daoguang Zan, Bei Chen, Yongshun Gong, Junzhi Cao, Fengji Zhang, Bingchao Wu, Bei Guan, Yilong Yin, and Yongji Wang.
\newblock Private-library-oriented code generation with large language models.
\newblock {\em arXiv preprint arXiv:2307.15370}, 2023.

\bibitem{zeng2025glm}
Aohan Zeng, Xin Lv, Qinkai Zheng, Zhenyu Hou, Bin Chen, Chengxing Xie, Cunxiang Wang, Da~Yin, Hao Zeng, Jiajie Zhang, et~al.
\newblock Glm-4.5: Agentic, reasoning, and coding (arc) foundation models.
\newblock {\em arXiv preprint arXiv:2508.06471}, 2025.

\bibitem{zhang2025towards}
Ziyue Zhang, Xianjin Xia, Ruonan Li, and Yuanqing Zheng.
\newblock Towards next-generation global iot: Empowering massive connectivity with harmonious multi-network coexistence.
\newblock In {\em Proceedings of the ACM SIGCOMM 2025 Conference}, pages 1009--1024, 2025.

\bibitem{zhou2022docprompting}
Shuyan Zhou, Uri Alon, Frank~F Xu, Zhiruo Wang, Zhengbao Jiang, and Graham Neubig.
\newblock Docprompting: Generating code by retrieving the docs.
\newblock {\em arXiv preprint arXiv: 2207.05987}, 2022.

\end{thebibliography}

\appendix
\section{RFC Tree Construction}\label{appendix:rfc_tree}
Protocol specification documents such as RFCs typically include a well-structured table of contents and have clearly defined sections.
We first use regular expressions to clean the RFC document by removing irrelevant information such as headers and footers. 
Next, we traverse the beginning of the RFC line by line to extract metadata including the RFC number, title, abstract, and table of contents. 
Based on the extracted table of contents, we construct a hierarchical RFC section tree, where each node contains the section number, title, child nodes (subsections), and parent node (section).
Finally, we traverse the main body of the RFC, extracting the content of each section and populating the corresponding nodes in the section tree.

\section{Prompt Templates}

\subsection{Protocol Understanding}\label{appendix:low-level}

\begin{figure}[htbp]
    \centering
    \includegraphics[width=0.95\linewidth]{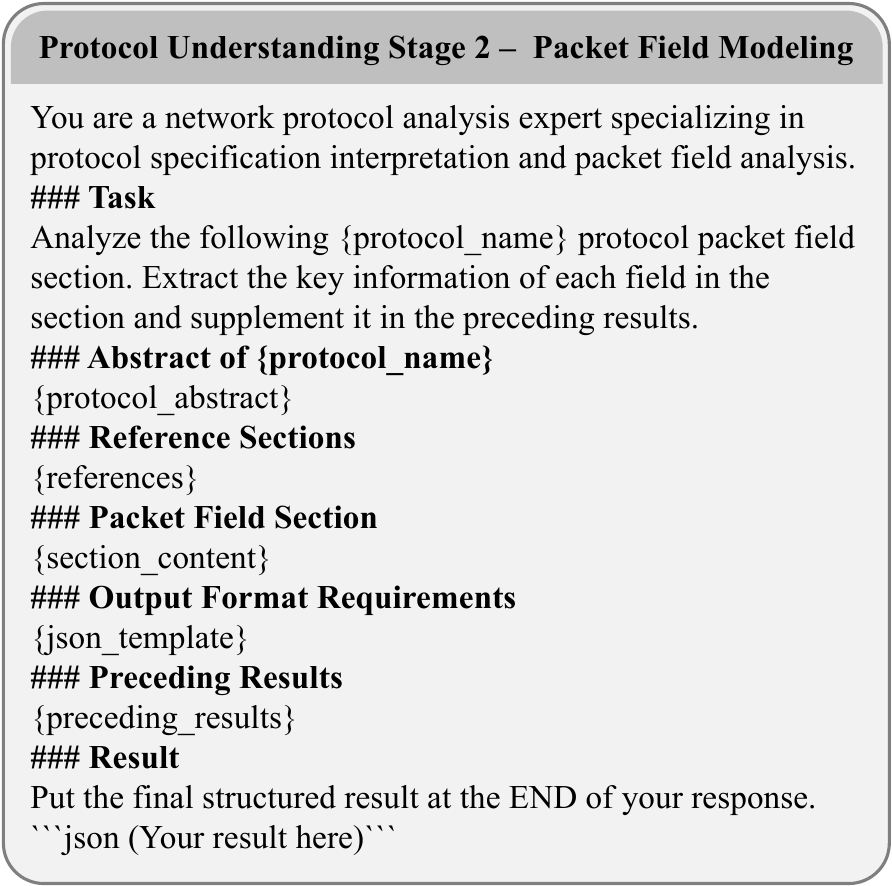}
    \caption{Prompt template of packet field modeling.}
    \label{fig:prompt-field}
\end{figure}

\S~\ref{subsubsec:high-level-analysis} present the prompt of the high-level analysis agent in Figure~\ref{fig:prompt-module}.
Figure~\ref{fig:prompt-field} takes packet field modeling as an example to illustrate the prompt of the low-level modeling agents.

\subsection{Test Case Generation}\label{appendix:tcg_template}

\begin{figure}[htbp]
    \centering
    \includegraphics[width=0.95\linewidth]{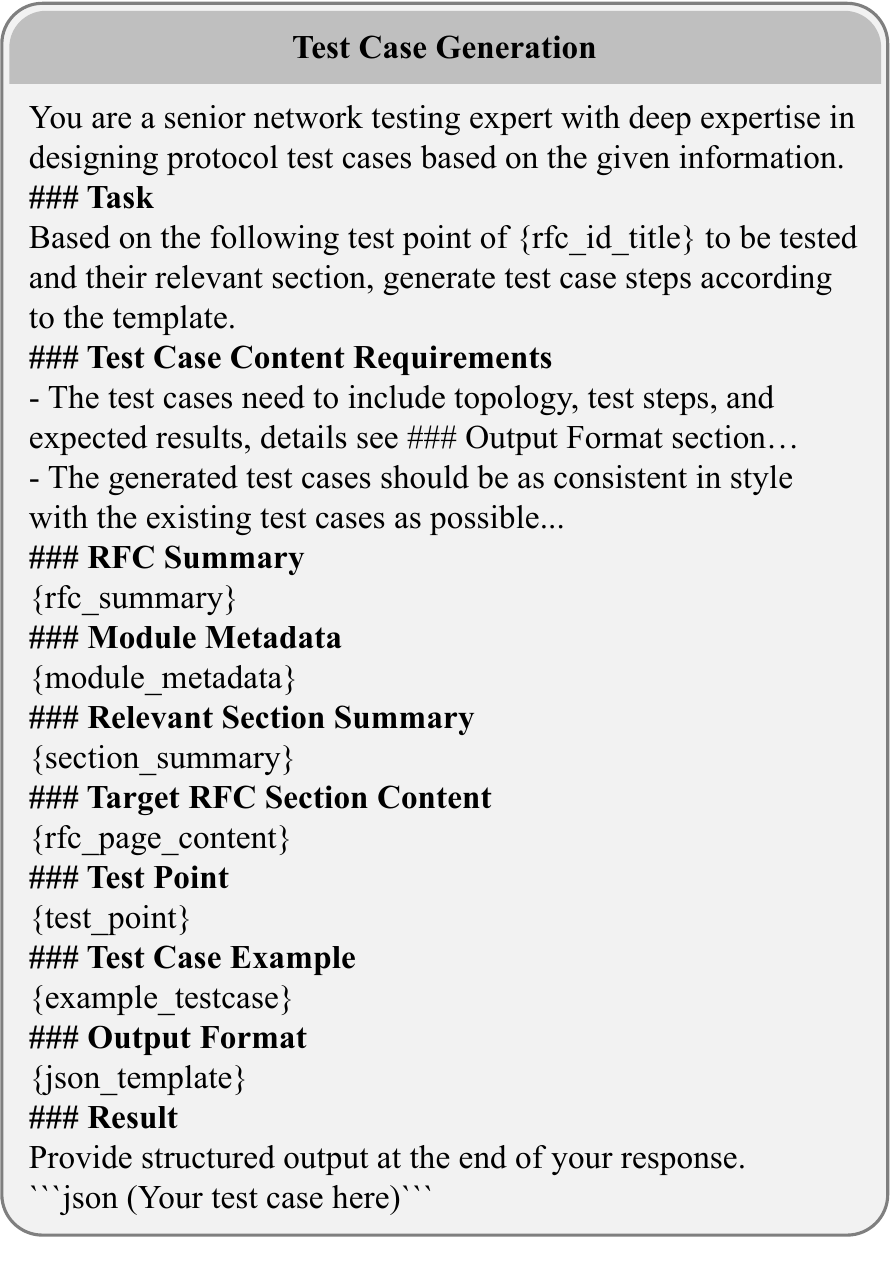}
    \vspace{-1ex}
    \caption{Prompt template of test case generation.}
    \label{fig:prompt-tcg}
\end{figure}

Figure~\ref{fig:prompt-tcg} shows the prompt template used for test case generation.
It guides the LLM to generate test cases based on the given test point, using the RFC summary, module metadata, relevant section overviews, and sample test cases as references.

\begin{figure}[htbp]
    \centering
    \includegraphics[width=0.95\linewidth]{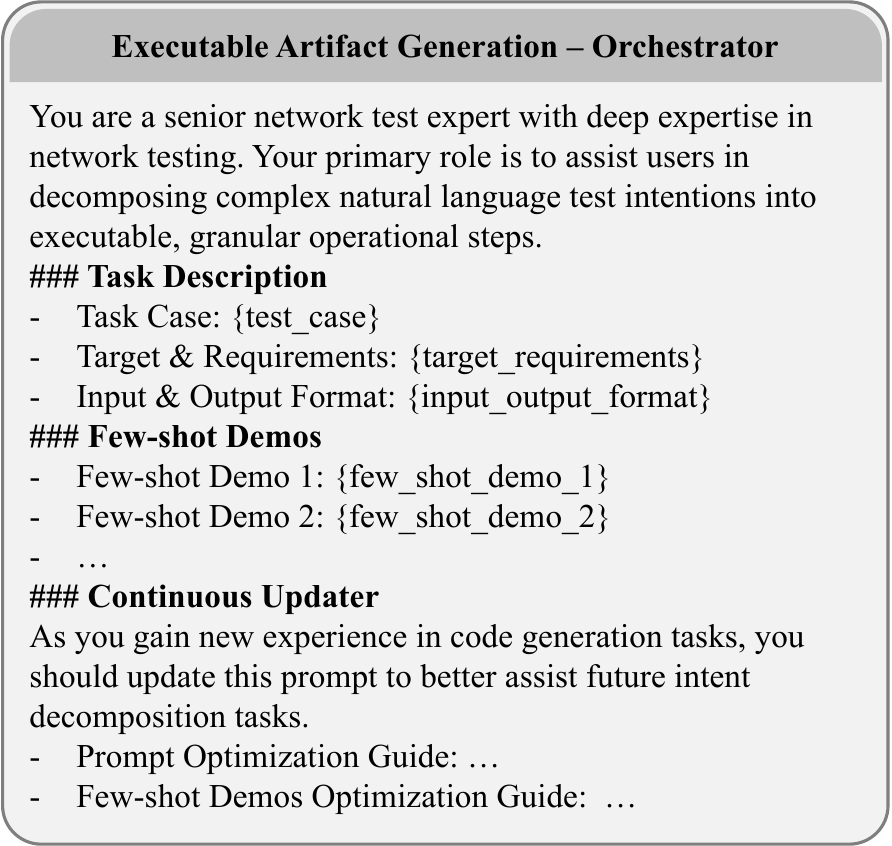}
    \caption{Prompt template of orchestrator.}
    \label{fig:prompt-tsg-orchestrator}
\end{figure}

\begin{figure}[htbp]
    \centering
    \includegraphics[width=0.95\linewidth]{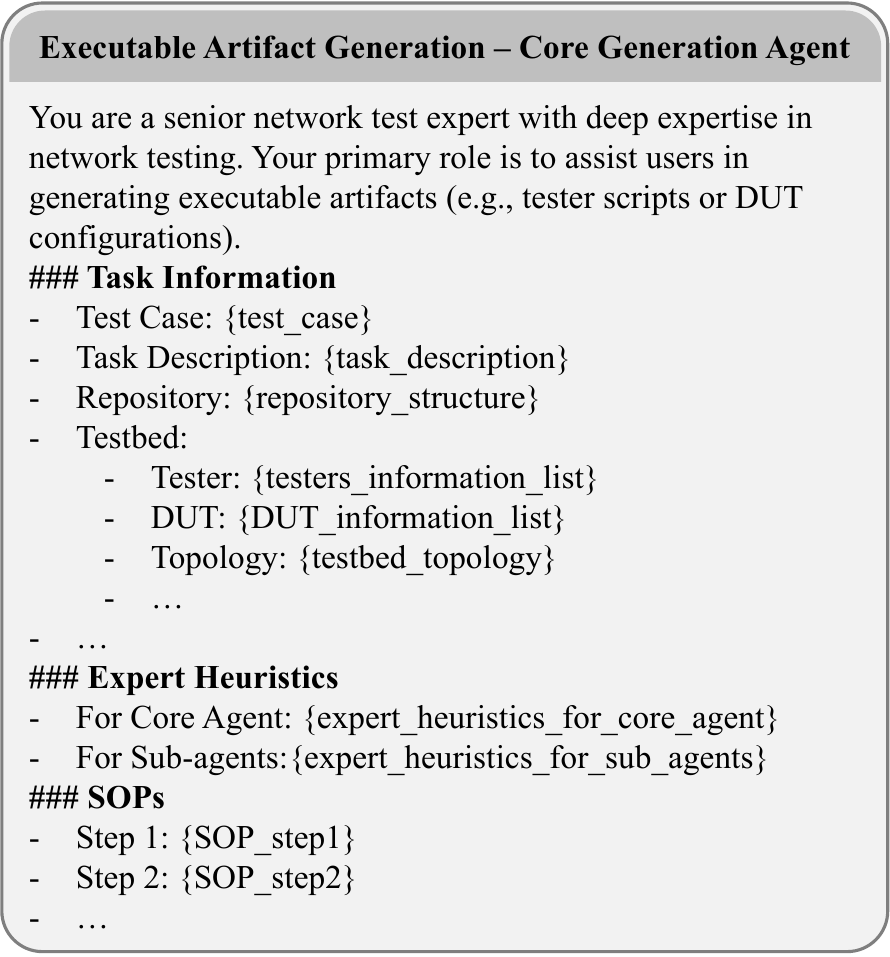}
    \caption{Prompt template of core generation agent.}
    \label{fig:prompt-tsg}
\end{figure}

\subsection{Executable Artifact Generation}\label{appendix:tsg_template}

Figure~\ref{fig:prompt-tsg} demonstrates the prompt template required by the core generation agent during the executable artifact generation process, which includes essential knowledge for completing the generation task, including (1) task information, (2) expert heuristics, and (3) SOPs, as described in \S~\ref{subsec:task-knowledge-base}.

As an example, Figure~\ref{fig:prompt-tsg-orchestrator} also demonstrates the prompt template for one of the sub-agents, the orchestrator, which includes: (1) \textit{task description}, comprising test case, task target and requirements, input and output format; (2) \textit{few-shot demonstrations}, containing a curated set of optimized few-shot examples; (3) \textit{continuous updater}, containing guidelines for self-updating its prompts or replacing few-shot examples.

\section{Survey Method of Expert Evaluation}\label{appendix:survey}
We categorize the 12 domain experts into three groups according to their domain expertise: junior, intermediate, and senior. 
For test case evaluation, we use the generated test case for OSPFv2 as the evaluation set and each expert is randomly assigned 20 test cases and asked to score them across four dimensions: correctness, completeness, testing value, and reproducibility. The final score of each test case is calculated as the average of these four dimensions.
For executable artifact generation, we provide each expert with five generated test scripts together with their corresponding test cases and reference scripts. Each expert evaluates the overall quality of the generated scripts.
In our scoring methodology, the total score is 10 points, with scores below 4 indicating \textit{not helpful}, 4-7 indicating \textit{helpful}, and scores above 7 indicating \textit{very helpful}.

\end{document}